\documentclass{emulateapj}
\usepackage{subfigure}
\usepackage{color}
\usepackage[colorlinks,linkcolor=blue,citecolor=blue,
urlcolor=magenta]{hyperref}

\shorttitle{Universality of the stellar mass--stellar metallicity
relation}
\shortauthors{Xia \& Yu}

\def\be{\begin{equation}}
\def\ee{\end{equation}}

\def\kms{{\rm \,km\,s^{-1}}}

\def\Mpc{{\rm \,Mpc}}

\def\msun{{\,M_\odot}}
\def\hot{{_{\rm hot}}}

\def\disk{{_{\rm disk}}}
\def\reheat{{_{\rm reheat}}}

\def\SN{_{\rm SN}}
\def\halo{{_{\rm halo}}}
\def\vir{_{\rm vir}}
\def\E{{_{\rm E}}}
\def\res{{_{\rm res}}}

\begin{document}
\newcommand{\feh}{[Fe/H]}
\newcommand{\mathfeh}{{\rm [Fe/H]}}

\title{Is the stellar mass--stellar metallicity
relation universal in the Milky Way satellites and beyond?}

\author{Moran Xia, \& Qingjuan Yu$^{\dagger}$}

\affil{Kavli Institute for Astronomy and Astrophysics, and School of Physics, Peking University, Beijing 100871, China}
\altaffiltext{}{$^\dagger$yuqj@pku.edu.cn}


\begin{abstract}

\noindent

Observations reveal a universal stellar mass--stellar metallicity relation
(MZR) existing in Local Group dwarfs of different types, $Z_*\propto
M_*^{\alpha}$ with $\alpha=0.30\pm0.02$. In this work, we investigate the
``universality'' of the MZRs for both satellites and central galaxies in a
large number of different host dark matter halos covering a large mass range of
$10^9$--$10^{15}h^{-1}\msun$, by using a semianalytical galaxy formation and
evolution model.  We obtain the following results. (1) The exponents ($\alpha$)
for the MZRs of the satellites in halos with the same mass as the Milky Way
halo but different individual assembly histories are mostly
$\sim$0.2--0.4, i.e., having a scatter of $\sim 0.2$; and the scatter of
$\alpha$ increases with decreasing halo masses. (2) The MZR relations are
changed little by the variation of halo masses and the classification between
central galaxies and satellites, if many halos with the same mass are stacked
together. (3) A double power law exists in the MZR relations for both central
galaxies and stacked satellites, with $\alpha\sim$0.2--0.4 at $10^3\msun\la
M_*\la10^{8}\msun$ and a relatively higher $\alpha\sim0.5$ at $10^8\msun\la
M_*\la10^{11}\msun$.  (4) The high-mass satellites ($M_*\ga10^8\msun$) existing
mostly in high-mass halos can lead to an apparent increase of $\alpha$ (from
$\sim0.2$ to $\sim0.4$) with increasing host halo masses shown in the
single power law fitting results of stacked satellites.  The universality of
the MZR suggests the common physical processes in stellar formation and
chemical evolution of galaxies can be unified over a large range of galaxy
masses and halo masses.

\end{abstract}

\section{Introduction}
\label{sec:intro}

The dwarf satellites in the Milky Way (MW) are potentially powerful probes of
the baryonic processes in galaxy formation occurring in the early universe.
Observations reveal that there exists a universal stellar mass--stellar
metallicity relation in the MW/M31 dwarf satellites and some other dwarf
irregular galaxies in the Local Group, $Z_*\propto M_*^\alpha$ with
$\alpha=0.30\pm0.02$ (\citealt{Kirby13}, K13 hereafter). The correlation
between the stellar metallicity and the stellar mass was reproduced in the
satellites of MW-like galaxies by using the semianalytical galaxy formation and
evolution model (e.g., \citealt{font2011,Lietal10,Luetal14}; \citealt{HYL14},
HYL14 hereafter).  In this work, we investigate how universal the stellar
mass--stellar metallicity relation (MZR) obtained from the semianalytical model
is, that is, whether the MZRs obtained for the following different cases could
fall onto the same relation: a) same halo masses but with different halo
assembly histories; b) different halo masses; c) satellite galaxies or central
galaxies.  Here by ``universal'' we mean not only the ``universal'' existence
of a correlation between the stellar metallicity and the stellar mass, but also
the ``universal'' or roughly the same quantitative ranges for the slope
($\alpha$) and the normalization of the correlation.  The exploration of
whether this relation exists and how universal it is will help us to understand
the common physical processes (e.g., star formation, chemical evolution)
involved in the formation and evolution of different galaxies, to reveal the
origin in shaping the relation, and to further provide constraints on galaxy
formation and evolution models.

In this work, we employ dark matter halo merger trees and a semianalytical
galaxy formation model (\citealt{Coleetal00}; see also
\citealt{WhiteFrenk91,Kauffmann93,Somerville99,Somervilleetal08}) to generate galaxies and their
satellites, as done in HYL14. The DM halo merger trees are generated from the
modified extended Press-Schechter function
\citep{PS74,Bondetal91,LC93,Somervilleetal08} by the Monte Carlo method.  The
model provides an efficient way to explore the effects of different physical
processes on observational properties. This exploration can be done by varying
the model parameters of the physical processes and allows a statistical study
by generating a large number of dark matter merger trees for each set of
parameters. HYL14 uses this method to obtain the MZR for the dwarf satellites
of MW-like central galaxies in MW-size dark matter halos, which is consistent
with the observation of MW satellites; and the slope in the MZR is shown to be
affected by the SN feedback strength and the reionization epoch.  Note that in
the semianalytical model not every central galaxy hosted by an MW-size dark
matter halo is MW-like. Also note that the correlation between stellar
metallicities and stellar masses exists not only in Local Group dwarfs, but
also in local galaxies with stellar mass $10^9\la M_*/M_{\odot}\la 10^{12}$ as
revealed by the SDSS spectra of over 40,000 galaxies
(\citealt{Gallazzietal05,Gallazzietal06}; K13; see also
\citealt{Panteretal08,GDetal14}). In this work, we generalize our study to
non-MW-like galaxies and to a large range of host halo masses
($\sim10^9\msun$--$10^{15}\msun$). We explore whether the MZR is universal in
the central galaxies with a large range of stellar masses ($\sim
10^3\msun$--$10^{12}\msun$) and in their satellites.

The relation between the metallicities of galaxies and their masses resulting
from semianalytical galaxy formation models has been explored extensively in
the past (e.g. \citealt{Y13,Somervilleetal15,Luetal17,font2011,guo11,Lietal10,Luetal14};
HYL14). Our
study is distinguished from previous works by a few aspects of the scope, the
purpose, and the method, as follows.
\begin{itemize}
\item As mentioned above, the stellar mass and the halo mass of the MZR studied
in this paper cover large ranges (with stellar masses $\sim
10^3$--$10^{12}\msun$ and halo masses $\sim 10^9$--$10^{15}\msun$), and the
galaxies include both central galaxies and their satellites. The mass ranges
covered in previous studies are smaller, and many do not put central galaxies
and their satellites together in the study.  For example, \citet{Y13}
obtain the stellar mass--cold gas metallicity relation for central galaxies
with relatively high stellar mass range $\sim 10^9$--$10^{11}\msun$.
\citet{Somervilleetal15} present the stellar mass--cold gas metallicity
relation for central galaxies with stellar mass range $\sim
10^7$--$10^{11}\msun$. \citet{Luetal17} present the MZR of MW dwarfs
with stellar mass range $\sim 10^3$--$10^9\msun$.
\item The previous studies do not investigate how the MZRs are different in a
wide range of galaxy masses and halo masses, and they focus mainly on
discussing different aspects or the effects/roles of different recipes/physical
processes in semianalytical models.  For example, \citet{Y13} discuss
the effects on the element abundances obtained by updating the chemical
evolution model and including delayed enrichment from stellar winds, SNe II and
SNe Ia, etc., in local star forming galaxies, elliptical galaxies, and MW-like
disk galaxies. \citet{Somervilleetal15} discuss the effects on the evolution
of some fundamental galaxy properties (e.g., stellar mass functions, the
relation between stellar mass and star formation rate, the relation between
stellar mass and cold gas phase metallicity) obtained by partitioning cold gas
into different phases and modeling the conversion of molecular gas into stars.
\citet{Luetal17} discuss the importance of different types of feedback
mechanisms (including both ``preventive feedback'' and ``ejective feedback'')
to explain the observational properties of MW satellites.  \citet{font2011}
illustrate that chemical properties can be used to break the degeneracy in
the effects of SN feedback and reionization on the luminosity function of MW
dwarf satellites.  As mentioned above, the physical properties obtained in
those studies do not cover a wide range of galaxy masses and halo masses as
done in this paper.
\item We adopt the Monte Carlo method based on the modified extended
Press-Schechter function to generate a large number (e.g., 100 or more trees)
of halo merger trees, which provides an efficient way to perform a statistical
study of the MZR. By contrast, the number of the merger trees obtained with
sufficiently high resolutions from cosmological N-body simulations are quite
limited, e.g., six MW-size halo assembly examples provided by the
high-resolution Aquarius dark matter simulations(\citealt{Springel08,S13};
see also the Via Lactea simulation, \citealt{Diemand07,Rocha12}).
\end{itemize}

Different scenarios have been proposed to explain the correlation between the
metallicities of galaxies and their masses (e.g., K13;
\citealt{Brooksetal07,Koppenetal07,D07,Lillyetal13,Maetal16,FD08,Luetal17,DS86},
and a review in \citealt{F16} and references therein). The study of
semi-analytical models provides a way to reveal the common reasons that lead to
a universal relation. In this paper, we mainly show the MZR obtained from
semi-analytical models and illustrate the universality.  In our next paper,
with the tool of the semianalytical model and its results, we present our
detailed chemical evolution model and the explanation for the universal MZR
revealed in this paper.

This paper is organized as follows. In Section 2, we briefly describe the
semi-analytical galaxy formation and evolution model used in this work.
Previously we have shown that the MZR in the satellites of MW-like galaxies
produced from the model matches the observation, given a specific set of
parameters (i.e., the fiducial model parameters in HYL14), such as on the
physical processes of feedback and the reionization of the universe. Then we
fix these parameters and explore the variations as a result of changing halo
masses, central galaxies, and dwarf satellites in Section 3. Remarkably, we
find that the MZR is quite independent of changing these variations as long as
the fiducial model parameters are fixed. A summary is given in Section 4.

In this paper we set the Hubble constant as $H_0=100\,h\kms\Mpc$, and the
cosmological model used is $(\Omega_{\rm
m},\Omega_\Lambda,h,\sigma_8)=(0.25,0.75,0.70,0.90)$.

\section{Method}
\label{sec:method}

In this section, we briefly describe the semianalytical galaxy formation model
that is used to explore the MZR of MW dwarf satellites and beyond in this work.
The backbone of the model is the merger trees of DM halos, which may represent
the hierarchical growth history of the host halo. Detailed semianalytical
recipes for galaxy formation and evolution (for reference, see
\citealt{Coleetal00,WhiteFrenk91,Kauffmann93,Somerville99}) are incorporated
into the merger trees to obtain the observational properties of the central
galaxy and its satellites.

We plant the halo merger trees using the Monte Carlo method developed by
Parkinson et al.\ (2008; see also \citealt{Kauffmann93,SK99,Coleetal00}), which
is based on a modified version of the extended Press-Schechter formula. The
obtained halo mass functions are in good agreement with the $N$-body simulation
results. The merger trees are built from redshift $z = 0$ to 20, with 79 equal
intervals in the logarithm of 1 + $z$.
The minimum progenitor halo mass set in the halo
merger trees is defined as the mass resolution $M\res$,
and the progenitors with mass lower than $M\res$ cumulatively
contribute to the accreted halo mass in the growth history of a halo (e.g., see
eq.~3.5 in \citealt{Coleetal00}).
For each halo mass, we generate a number of merger trees (e.g., 10000, 1000, or 10) and
apply the semi-analytical galaxy formation and evolution recipes to them.  The
related recipes, e.g., on SN feedback, reionization, and gas cooling, are
summarized below in this section.  More details can be found in HYL14 and
references therein.  In the assembly history of a halo, a satellite of the
present-day central galaxy was the host galaxy of a small halo at an early time
before it fell into a big halo. 

As described in HYL14, the semianalytical galaxy formation model in this study
is based on GALFORM \citep{Coleetal00,galform2,Boweretal06}, but with several
modifications.  As demonstrated by \citet{Coleetal00}, \citet{Kauffmann93},
\citet{Somerville99}, \citet{Croton06}, and \citet{Boweretal06}, the
semianalytical models can successfully reproduce a number of observations on
the statistical distributions of galaxy properties, including the galaxy
luminosity function, the stellar mass function, etc., and different
observational constraints can provide probes of different underlying physical
processes.  \citet{Somervilleetal15} show that both the MZR and the stellar
mass function at $z=0$ in the stellar mass range of $10^7$--$10^{11}\msun$ can
be reproduced.  \citet{Luetal17} discuss the status in reproducing the MZR
simultaneously with other observational constraints in the stellar mass range
of $10^3-10^9\msun$, and suggest that both ejective feedback (strong outflow)
and preventive feedback (photoionization heating in low-mass halos) are
necessary to reproduce the MZR and the stellar mass function. As shown by
Figures 1 and 3 in HYL14, our model has reproduced both the luminosity function
and the MZR in MW-like halos, with including the effects of both reionization
and SN feedback.  In this study, we focus on generalizing the same
semianalytical model to see how the MZRs vary in a large range of different
halo masses and galaxy masses.

Specifically, to study the properties of the MW satellites, MW-size halos with
mass $\sim$1--2$\times 10^{12}h^{-1}\msun$ are set in the model and MW-like
galaxies are selected from the generated galaxies, as done in HYL14.  The
selection criteria for the MW-like host galaxies is that the total stellar mass
of a present-day host galaxy is in the range of $4-6\times 10^{10}\msun$ and
its bulge mass to disk mass ratio is between 0.1 and 0.4 (e.g.,
\citealt{LN15,M11}). 

In this work, to address the universality of the MZR, we generalize the study
to galaxies and DM halos with different masses, so that the properties of
galaxies with different masses and their satellites in different environments
can be explored.

Below are the related semi-analytical galaxy formation and evolution recipes.

\begin{itemize}

\item Gas cooling:
In a newly formed dark matter halo, the initial total hot gas mass is a sum of
the total hot gas mass from its progenitor halos (with mass higher than $M_{\rm
res}$) and the accreted gas mass.  In a halo merger tree, the accreted gas mass
is the baryon mass in the accreted progenitor halos with mass lower than
$M_{\rm res}$. The baryon fraction in the accreted halo mass 
is obtained by the cosmic average baryon fraction multiplied by a
fraction reduced by the reionization of the universe. The fraction reduced by
the reionization of the universe will be described below (or see eq.~1 in
HYL14). The effect of reionization is also included in the estimate of the
total hot gas mass that can be kept from its progenitor halos with mass higher
than $M_{\rm res}$.  The initial temperature of the hot gas in the newly formed
halo is set to be the virial temperature of the halo.
 
The hot gas in the dark matter halo cools down to the halo center through
atomic cooling with a rate depending on the temperature, mass density, and
metallicity of the hot gas \citep{SD1993}. The cooling recipe of the hot halo
gas in this work is adopted from HYL14, which is based on the GALFORM model and
modified by including the mixing of the gas reheated by SN feedback with hot
halo gas (mainly occurring in low-mass galaxies or progenitors of dwarf
satellites).

As done in HYL14, the molecular hydrogen cooling process for pristine gas that
occurred in the mini-halos in the early universe is modeled by using Equations
(21)-(29) in Benson (2010; see also \citealt{galli1998}), and there is no
molecular hydrogen cooling after the completeness of the reionization of the
universe due to the suppression of the abundance of hydrogen molecules caused
by the strong UV background (e.g., \citealt{W17}). HYL14 shows that the MZR of
the satellites in MW-like galaxies is not sensitive to whether the molecular
hydrogen cooling recipe is included or not.

\item Star formation and stellar feedback: following \citet{Coleetal00}, stars
are formed in the disk at a rate directly proportional to the mass of cold gas
in the disk, given by
\begin{eqnarray}
&& \psi =M_{\rm cold}/\tau _* \\
&& \tau _*=\epsilon _*^{-1}\tau\disk(V\disk/200\kms)^{\alpha _*},
\label{eq:sim_sf}
\end{eqnarray}
\noindent
where $\psi$ is the instantaneous star formation rate, $\tau_*$ is the star
formation time scale, $\tau\disk$ is the dynamic time scale of the galaxy
disk, $V\disk$ is the disk rotation velocity, and $\epsilon_*=0.005$ and
$\alpha_*=-1.5$ are two parameters (see eq.\ 4.14 in \citealt{Coleetal00}).

Star formation will re-heat the cold gas in the disk and possibly expel it out
of the galaxy. The gas mass reheated by the feedback during time interval $dt$
is given by:
\begin{eqnarray}
&& dM\reheat=\beta \psi dt,
\label{eq:dMreheat}
\\
&& \beta =(V\disk/V\hot)^{-\alpha\hot},
\label{eq:sim_fb}
\end{eqnarray}
where $\beta$ is the feedback efficiency, and $V\hot$ and $\alpha\hot$ are the
two parameters defining the strength of the feedback with $V\hot=200\kms$
and $\alpha\hot=3.2$.
Part or all of the reheated gas can escape the dark matter halo if the strength
of the total feedback energy released and coupling to the intergalactic medium (IGM) is sufficiently
large with
\begin{equation} dE\SN-\frac{1}{2}v\vir^2dM\reheat\geq 0,
\label{eq:ener_condi} \end{equation} where
$dE\SN=\epsilon\halo\times\frac{1}{2}v\SN^2\psi dt$ is the total energy
released by SNe and coupling to the IGM during time $dt$, $v\vir$ is the virial
velocity of the halo, $\frac{1}{2}v\SN^2$ is the total energy released per unit
mass by SNe with $v\SN=630\kms$, and $\epsilon\halo=0.05$ is the fraction of the energy that couples to
the cold gas in the disk.
For low-mass galaxies or the progenitors of satellites (e.g., $M_*\la 10^8\msun$), normally the
above inequality (\ref{eq:ener_condi}) cannot be satisfied, the reheated gas
that is expelled
out of the disk stays in the halo, with mass during time interval $dt$ given by
$\beta\E\psi dt$, where
\begin{equation} \beta\E\equiv dE\SN/(\frac{1}{2}v^2\vir\psi dt)=\epsilon\halo (v\vir/v\SN)^{-2} \label{eq:betaE} \end{equation}
(see more details in section 2 in HYL14).

The feedback due to SN Ia explosions is included in this work. We assume that
the energy released by an SN II explosion and an SN Ia explosion is the same.
We use the same feedback recipe for each
(Equations~\ref{eq:dMreheat}--\ref{eq:betaE}), but include the non-negligible
time delay between the formation of SN Ia progenitors and the SN Ia explosions
(See Equations 13-16 in HYL14).

As done in HYL14, in the assembly history of a halo, a satellite of the
present-day central galaxy may be the host galaxy of a small isolated halo
before it fell into a big halo at an early time. We apply the above energy
condition only to a galaxy before it becomes a satellite. We do not apply it to
satellites, but assume that the reheated gas from satellites (with mass
expected by Equation~\ref{eq:sim_fb}) is expelled into the big host halo, as
the original halos of the satellites are largely tidally disrupted along their
motion in the big host halo, and the tidal field induced by the big host halo
also helps to keep those expelled materials out of the satellites.  Note that
the tidal stripping and disruption of the stellar and cold gas components of
the satellites are not considered in our model, as they are located in a
smaller central region compared with their original halo size.  In \citet{S13},
the tidal stripping and disruption of satellites was shown not to have a
significant effect on the satellite luminosity functions, as it affects very
few satellites.

\item Metallicity enrichment: the metals ejected by SNe are assumed to be
homogeneously and instantaneously mixed with the interstellar medium in the
galaxy, and after the mixture, some metals can be ejected out of the galaxy
along with the mixed interstellar medium that is ejected out by SN explosions.
In this work, the Fe yield of SNe II is adopted from tables 2–3 in
\citet{SN_II_pattern}, and the Fe yield of SNe Ia is from
\citet{SN_Ia_pattern}.

\item The reionization of the universe: the reionization in the early universe
reduces the baryon fraction of a DM halo. The extent of the reduction is
modeled through a mass scale called the filtering mass, which is a function of
redshift, as well as a function of the redshift that the first ionized bubble
formed $z_0$, and the completion redshift of the reionization $z_r$.  A halo
with mass lower than the filtering mass loses more than 50\% of the baryonic
matter expected from the cosmic average. In this paper, we use Equations (B1)
and (B2) in Kravtsov et al.\ (2004; see also \citealt{gnedin2000,Okamoto08}) to
calculate the filtering mass and model the effects of the reionization.  In
this work, we set $z_0=15$ and $z_r=10$ as done in the fiducial model in HYL14,
which results in an MZR compatible with current observations of the MW
satellites. HYL14 shows that an early reionization epoch results in a
relatively high metallicity at the low-mass end and a relatively flat slope in
the MZR.  The relatively strong/early reionization epoch revealed in HYL14
suggests that the local universe is reionized earlier than the cosmic average;
local sources may have a significant contribution to the reionization in the
local region.

\end{itemize}

The parameters used in the recipes are chosen from the fiducial model in HYL14.

\section{Results}
\label{sec:res}

In this section, we show our simulation results of the MZR obtained by using
the method described in Section 2. Table~\ref{tab:runs} lists the simulation
runnings done in this work, with the host halo mass $M\halo$ ranging from $10^9
h^{-1}\msun$ to $10^{15}h^{-1}\msun$ at $z=0$ and the number of the merger
trees generated by the Monte Carlo method for each set of parameters. For low
halo masses, a mass resolution $M\res=10^4 h^{-1}\msun$ is set.
For high halo masses with $M\halo\ge 5\times 10^{11}h^{-1}\msun$, although the
values of $M\res$ are set to be relatively high, we set two sets of $M\res$ in
the runnings, which shows that the simulation results are close to being
convergent below. 

\begin{table}[t!]
\centering
\caption{Simulation Runnings}
\begin{tabular}{c|c|c||c|c|c}
\hline
\hline
$M\halo$ & $M\res$ & $N$ & $M\halo$ & $M\res$ & $N$\\
($h^{-1}\msun$) & ($h^{-1}\msun$) &  & ($h^{-1}\msun$) & ($h^{-1}M_{\odot}$) & \\
\hline
$10^9$ & $10^4$ & 10000 &  & & \\
$5\times 10^9$ & $10^4$ & 10000 & & &\\
$10^{10}$ & $10^4$ & 10000 &  &  & \\
$5\times 10^{10}$ & $10^4$ & 10000 & & &\\
$10^{11}$ & $10^4$ & 10000 &   &   &  \\
$5\times 10^{11} $ & $10^5$ & $1000$ & $5\times 10^{11}$ & $10^4$ & $10$\\
$10^{12} $ & $10^6$ & $1000$ & $10^{12} $ & $10^5$ & $10$\\
$2\times 10^{12} $ & $10^6$ & $1000$ & $2\times 10^{12} $ & $10^5$ & $10$\\
$5\times 10^{12} $ & $5\times10^6$ & $10$ & $5\times 10^{12} $ & $5\times10^5$ & $10$\\
$10^{13} $ & $10^7$ & $10$ & $10^{13} $ & $10^6$ & $10$\\
$10^{14} $ & $10^8$ & $10$ & $10^{14} $ & $10^7$ & $10$\\
$10^{15} $ & $10^9$ & $10$ & $10^{15} $ & $10^8$ & $10$\\
\hline
\end{tabular}
\tablecomments{The simulation runnings done in this work. $M\halo$ is the host
dark matter halo mass at redshift $z=0$, $M\res$ is the mass resolution of the
dark matter merger trees, and $N$ is the number of merger trees planted by
Monte Carlo runnings for each set of ($M\halo, M\res$). For $M\halo$ in the
range of $5\times 10^{11}$--$10^{15}h^{-1}\msun$, we set two different mass
resolutions in the runnings to check the convergence of the results, with
$M\halo/M\res\sim 10^6$ in the left column set and
$\sim 10^{7}$ in the right column set. }
\label{tab:runs}
\end{table}

\subsection{The MZR of the satellites in MW-size halos}
\label{subsec:MWlike}

\begin{figure*}[htb] \centering
  \subfigure{\includegraphics[scale=0.75]{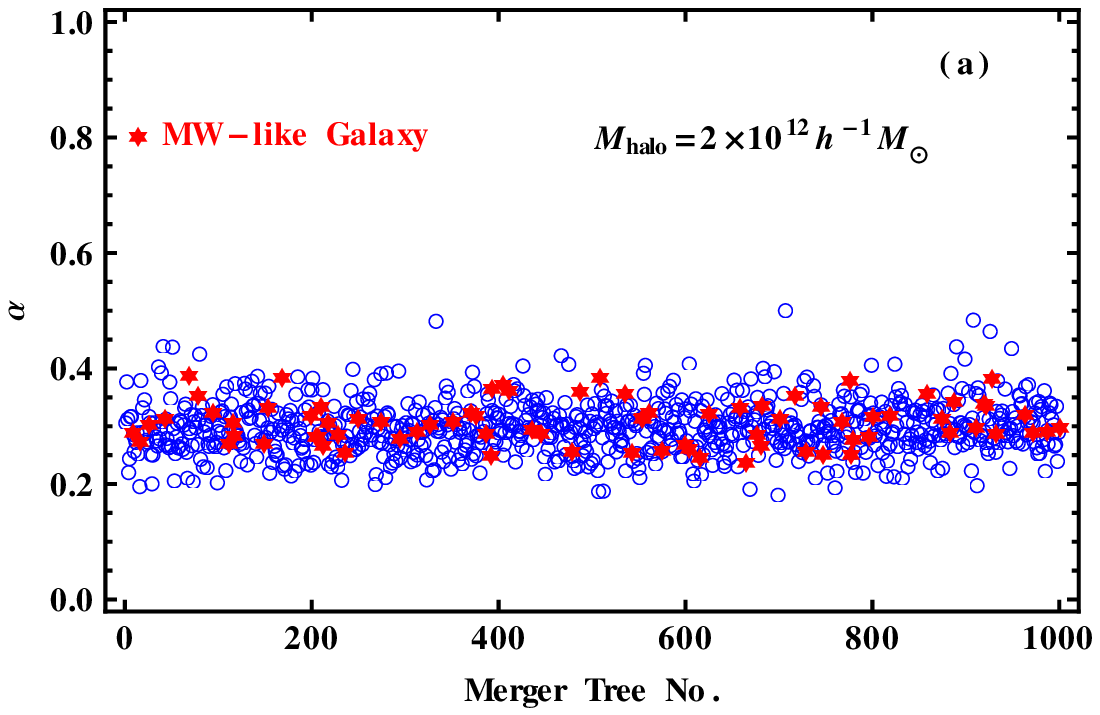}}
\subfigure{\includegraphics[scale=0.75]{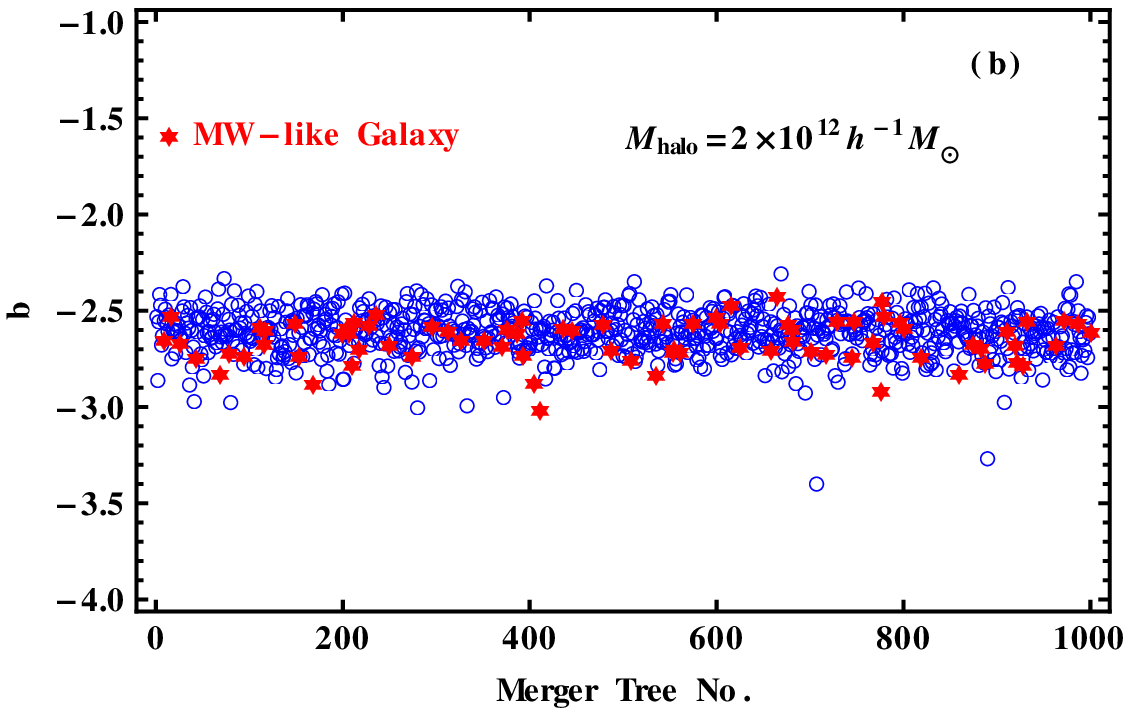}}
  \subfigure{\includegraphics[scale=0.75]{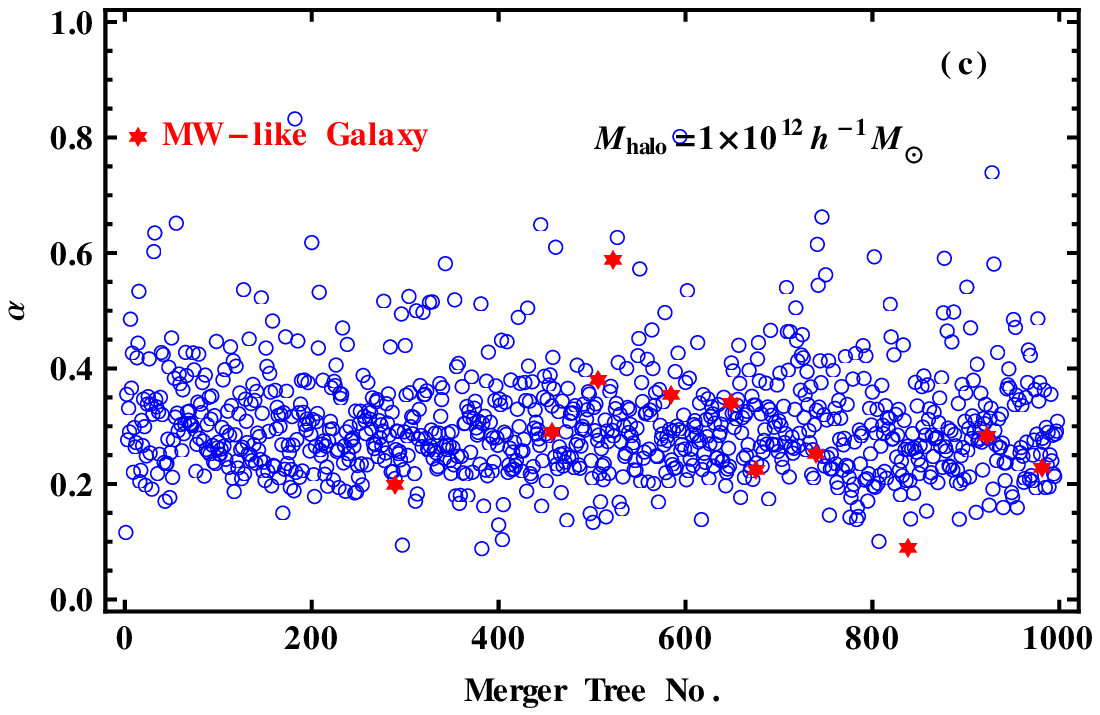}}
\subfigure{\includegraphics[scale=0.75]{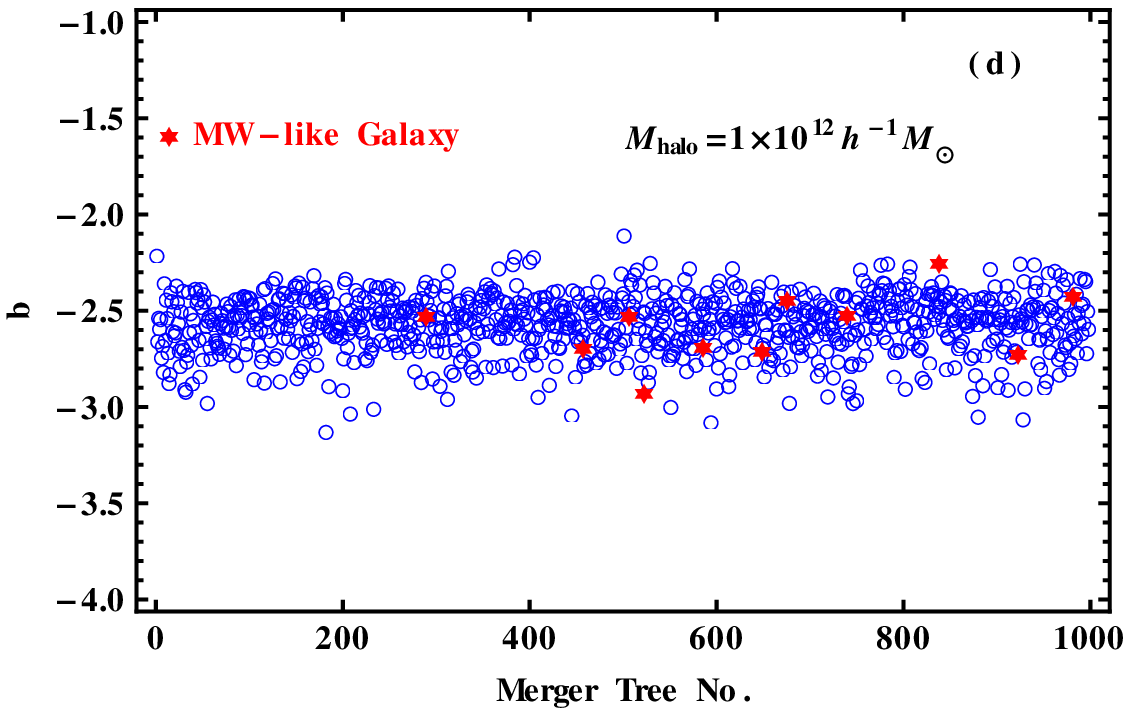}}
  \caption{The MZR linear fitting results for the simulated satellites in MW-size
dark matter halos (see Equation~\ref{eq:MZR}).  The host halo mass at redshift
$z=0$ is set to $2\times 10^{12}h^{-1}M_{\odot}$ in the top panels and
$10^{12}h^{-1}M_{\odot}$ in the bottom panels, with $M\res=10^6 h^{-1}\msun$.
Each panel shows the results of 1000 dark matter halos, and each point
represents the best-fit result to the satellites in one halo. For clarity,
the error bar of each point is not shown in the figure. The systems in which
the central galaxy is MW-like are marked with red stars, which do not appear
exceptional among all the systems.  This figure demonstrates that the MZR of the
satellites in MW-like halos is roughly universal with different individual
halos, and the best-fit slopes and
intercepts of the satellites of the halos are distributed mostly within the
same ranges, with $\alpha\sim$0.2--0.4 and $b\sim-2.9$--$-2.5$.  See details in Section~\ref{subsec:MWlike}.  }\label{fig:mw100}
\end{figure*}

The MW halo mass is in the range $0.8-4.5\times 10^{12}\msun$
(e.g., \citealt{PND13,Kafleetal14,BK13,Callinghametal18,PH18,Watkinsetal18},
and references therein). As shown in Table~\ref{tab:runs}, we use
the Monte Carlo method to generate 1000 halo merger trees for both
$M\halo=10^{12}h^{-1}\msun$ and $M\halo=2\times10^{12}h^{-1}\msun$
at $z=0$ (with $M\res=10^6 h^{-1}\msun$)
and incorporate the baryonic processes described in Section 2 into the
halo merger trees to obtain the MZRs of the central galaxies and their
satellites in the halos.
An example of the MZR of the satellites in the MW-size halos
obtained from the simulations is illustrated by the dots in
Figure~\ref{fig:sats_full}(b)-(c).
The similar results are also shown in figure 3 in HYL14.

Figure~\ref{fig:mw100} shows the fitting results for the MZR of the satellites
in the 1000 host halos with mass $M\halo=10^{12}h^{-1}\msun$ and
$2\times10^{12}h^{-1}\msun$.
Note that the satellites are not uniformly distributed in the logarithm of the
stellar mass, and there are more low-mass systems than high-mass ones.  To
avoid that low-mass satellites contributing too much weight in the fitting
(which is more significant in high-mass halos), 
the fitting is performed in the following way. For
each halo, the logarithm of the stellar mass of the satellites of the host
galaxy at $z=0$ is divided into some bins starting from $\log (M_*/\msun)=3$
with an interval of 1 dex (see an example of stacked halos
in Figure~\ref{fig:sats_full}).
Noting that the stellar masses of the satellites are still not uniformly
distributed within a bin and we do not pre-assume that the stellar
metallicities follow a symmetric distribution or an exact Gaussian distribution
at a given stellar mass, we take the medians\footnote{In practice, we have
tested that our results
and conclusions will be changed little if we take the means instead
of the medians in each bin.} of the variables ($\log
M_*$,[Fe/H]) of all the satellites in each bin, and using the least-squares
method to fit the medians as follows,
\begin{equation}
{\rm [Fe/H]}=\alpha\log (M_*/10^3\msun)+b,
\label{eq:MZR}
\end{equation}
where $\alpha$ is the best-fit slope and $b$ is the best-fit intercept at $\log
(M_*/\msun)=3$. In the fitting, the error bars of the medians of [Fe/H] are
given by the half of the range between the 16th and 84th percentiles of the
distribution of the [Fe/H] of the satellites. 
Each point in Figure~\ref{fig:mw100} represents
the best-fit result to the satellites in one host halo at $z=0$. Those systems
in which the central galaxy is MW-like are marked in red, which are selected by
the criteria described in Section 2.  This figure demonstrates that the MZR of
the satellites in MW-size halos is roughly universal, with the slopes and the
intercepts of the satellites of the halos are distributed mostly within the
same range.  As seen from the figure, the fitted slopes $\alpha$ are
scattered mostly in the range 0.2-0.4, and the fitted intercepts $b$ are
scattered mostly in the range [-2.9,-2.5].  In the bottom panels with the
relatively low halo mass, the slopes have relatively large scatters with large
$\alpha$ up to 0.6, which is mainly because the number of the satellites are
relatively small in a relatively low-mass halo and the corresponding
statistical uncertainty is large. As seen from Figure~\ref{fig:mw100}, the MZRs of the
satellites are not sensitive to the selection criteria for their central
galaxies, as the red symbols representing MW-like central galaxies do not
appear exceptional among the total sample.

\subsection{The MZR of the satellites in a wide range of halo masses}
\label{subsec:multihalo}

\begin{figure*}[htb]
\centering
  \subfigure{\includegraphics[scale=0.55]{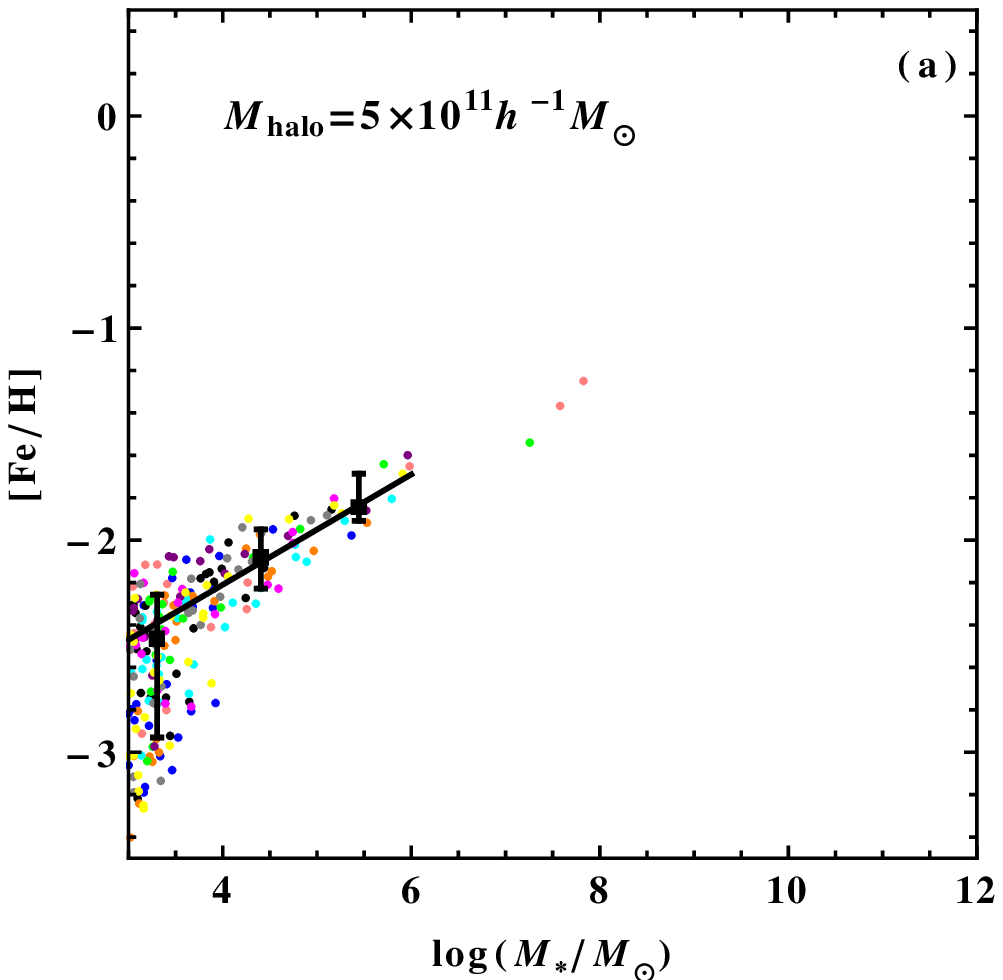}}
  \subfigure{\includegraphics[scale=0.55]{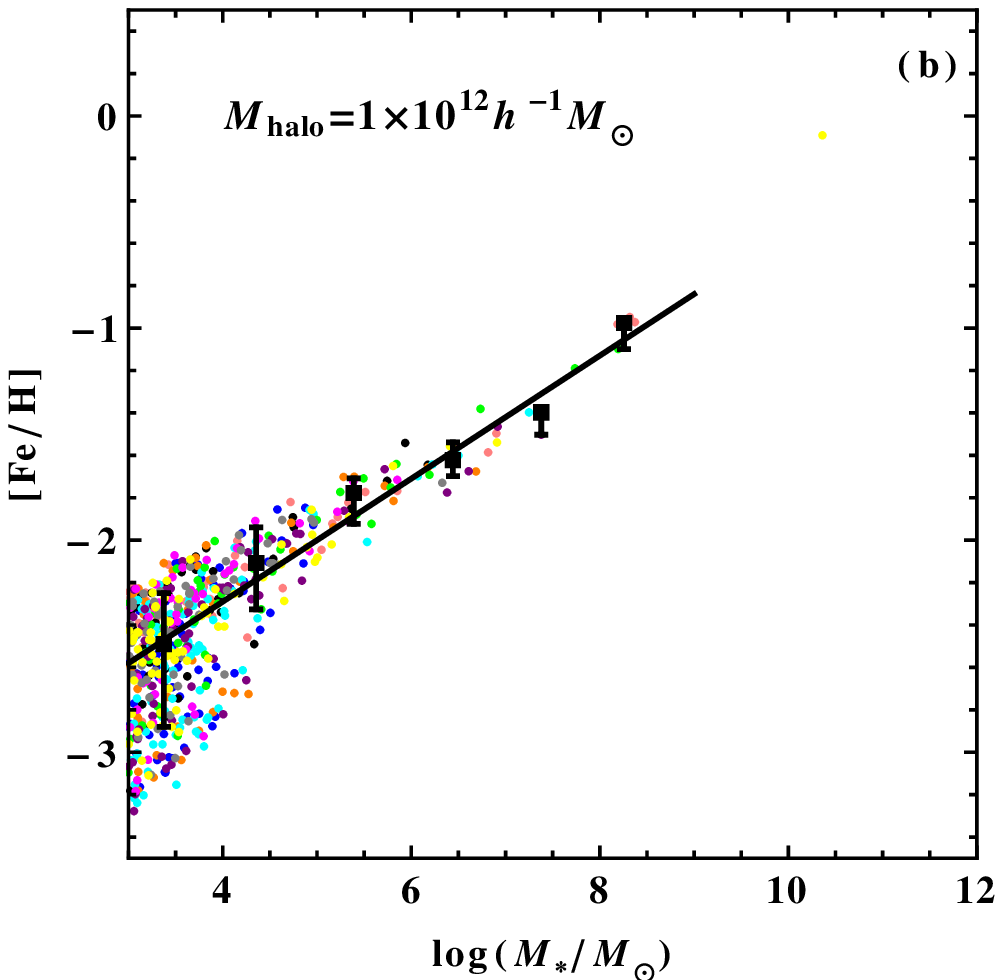}}
  \subfigure{\includegraphics[scale=0.55]{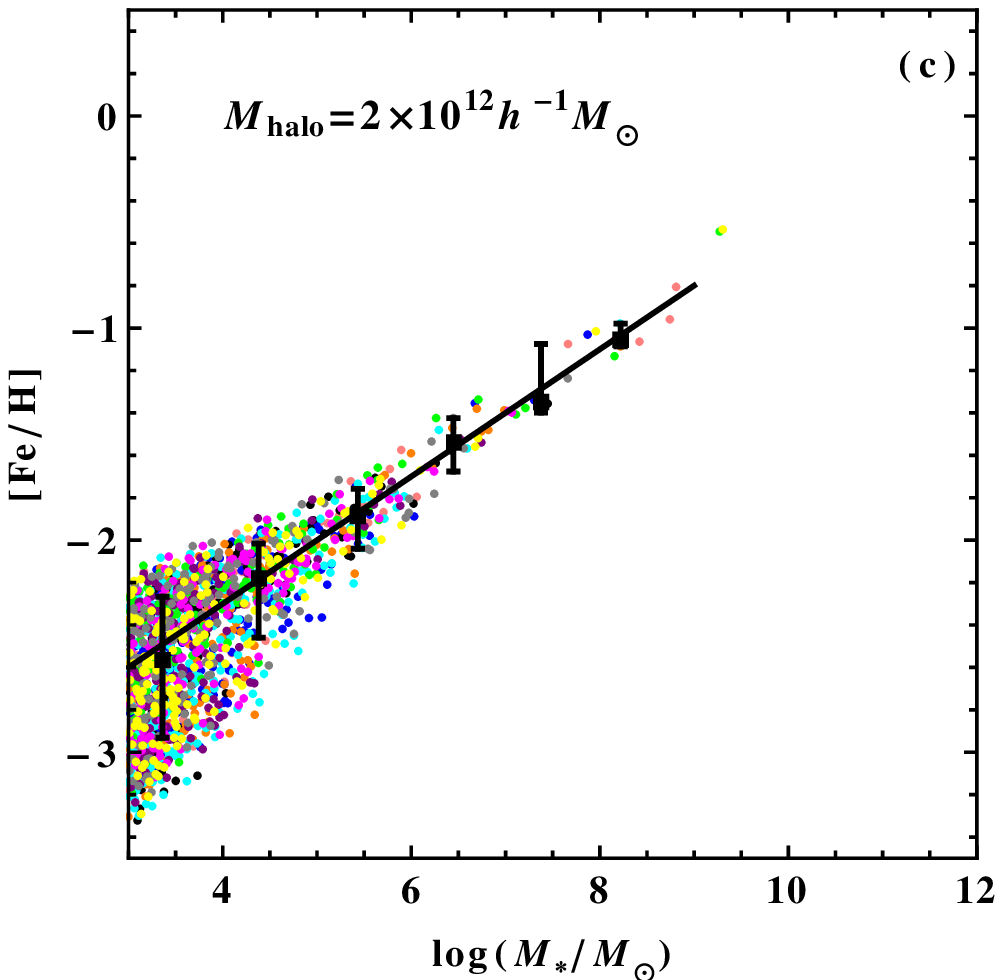}}
  \subfigure{\includegraphics[scale=0.55]{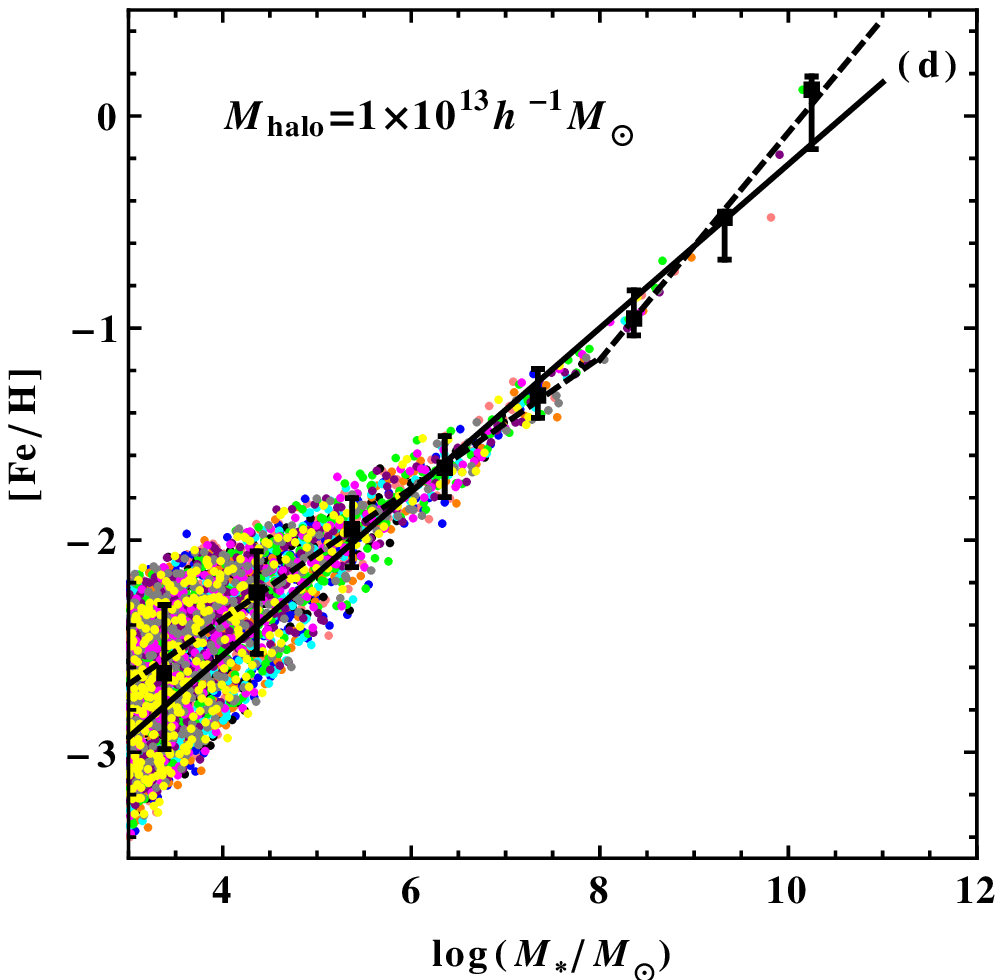}}
  \subfigure{\includegraphics[scale=0.55]{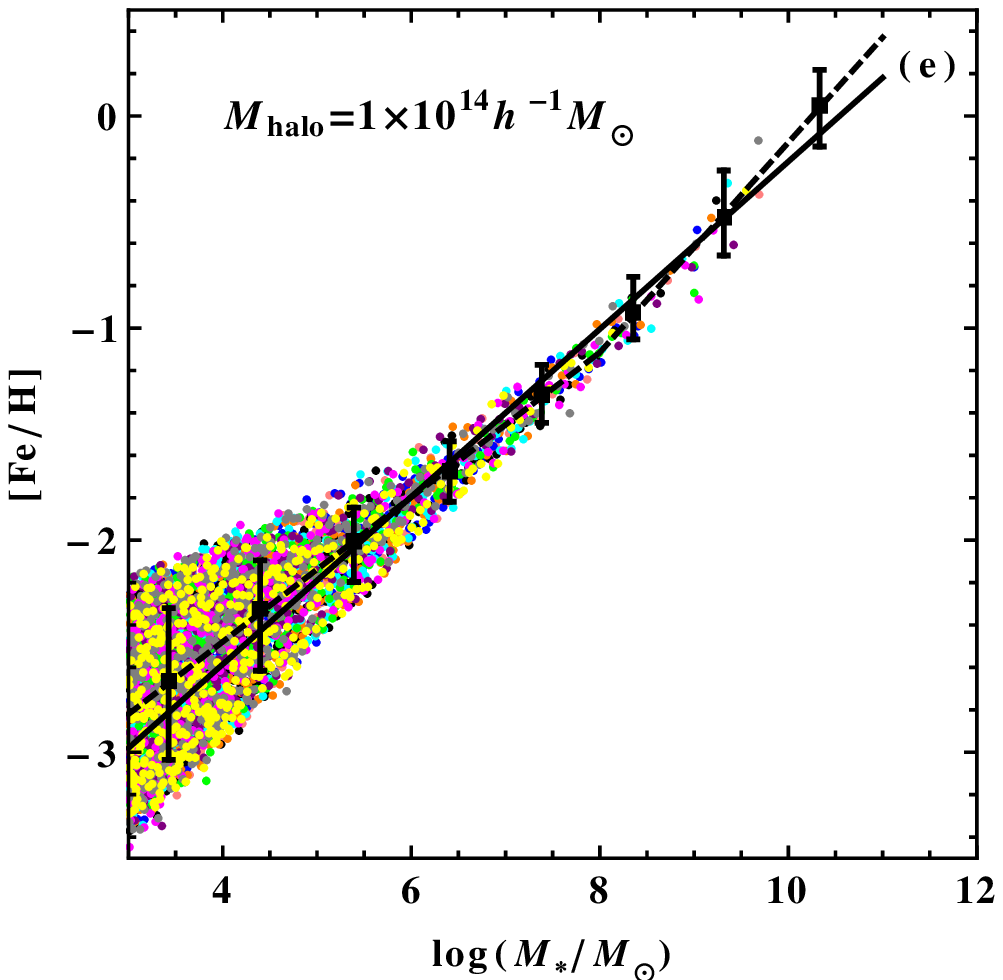}}
  \subfigure{\includegraphics[scale=0.55]{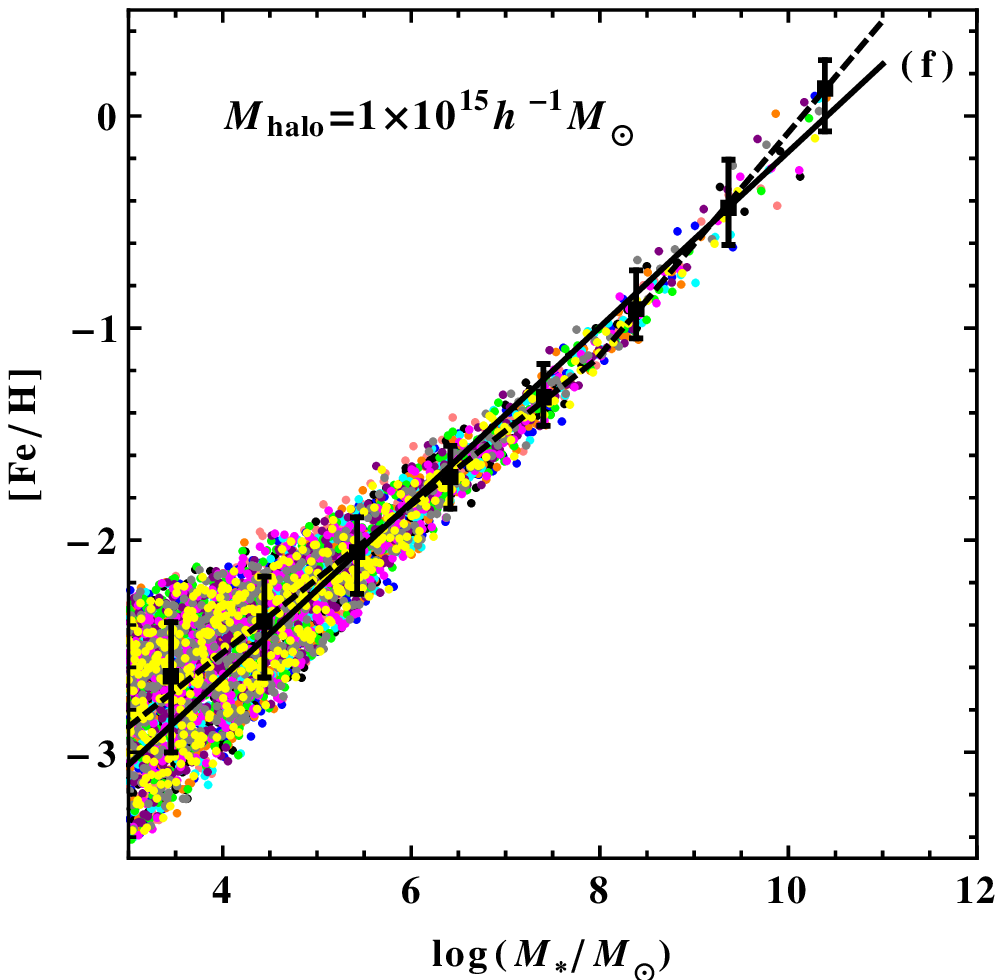}}
  \caption{The MZR for the simulated satellites in dark matter
halos with different halo masses. Each dot represents a satellite. Different panels show the results for
different host halo masses at $z=0$.  For the same host halo mass, the results
of 10 dark matter halo merger trees with $M\halo/M\res\sim 10^7$ are shown in
each panel, and different colors represent different trees. In the stellar mass
range shown in the figure, the average satellite number of each tree is 22, 57,
165, 1057, 10089, 62859 in panels (a)-(f), respectively. We show all the
satellites of the ten trees in the upper panels.  High-mass halos have too
numerous satellites, and for view clarity, we show only a fraction of them
randomly, i.e., 500 satellites for each tree in panels (d)-(f).  The
logarithm of the stellar mass is divided into some bins with a 1-dex interval,
starting from $\log(M_*/\msun)=3$, i.e., [3,4],[4,5],...; and in each bin the
black squares represent the medians of all the satellites with the stack of the 10
trees.  The error bars of the black squares represent the range between the
16th and 84th percentiles of the distribution of the dots in each bin.  The
linear (single power law) fitting to the black squares is shown by the black
solid line in each panel, and the dashed line in panels (d)-(f)
are the double power law fitting results.
The best-fit parameters are shown in Table~\ref{tab:stack} and 
Figure~\ref{fig:mzr_ab}(c)--(f).
This figure illustrates the similarity of the MZRs of stacked satellites in
different host halo masses. See Section~\ref{subsec:multihalo}.
}\label{fig:sats_full} \end{figure*}

\begin{table*}
\centering
\caption{The fitting results of the MZR}
\tiny{
\begin{tabular}{c|c|cccc|cccc|cccc}
\hline
\hline 
objects     & $M\halo$ 
& \multicolumn{4}{c|}{$N=10000$ or 1000\footnote{\tiny $N=10000$ for 
$M\halo\le 10^{11}h^{-1}\msun$and $N=1000$ for $M\halo= 5\times 10^{11}$--
$2\times 10^{12}h^{-1}\msun$, as listed in Table~\ref{tab:runs}.  } }
& \multicolumn{4}{c|}{$N=10$}  & \multicolumn{4}{c}{$N=10$}
\\

            & ($h^{-1}\msun$)
& \multicolumn{4}{c|}{ }
& \multicolumn{4}{c|}{$M\halo/M\res\sim 10^{7}$}  & \multicolumn{4}{c}{$M\halo/M\res\sim 10^{6}$} 
\\
\hline
  &  & $\alpha$ & $b$ & $\chi^2$ & $N_{\rm bin}$ & $\alpha$ & $b$
& $\chi^2$ & $N_{\rm bin}$ & $\alpha$ & $b$ & $\chi^2$ & $N_{\rm bin}$ \\

\hline 
& $5\times 10^{9}$ & $0.27\pm0.19$ & $-2.3\pm0.2$ & -  & 2 &  & & & & & & &\\
& $10^{10}$ & $0.24\pm0.09$ & $-2.3\pm0.2$ & $0.02$  & $3$  &  & & & & & & &\\
& $5\times 10^{10}$ & $0.25\pm0.04$ & $-2.4\pm0.1$ & $0.8$  & $5$  &  & & & & & & &\\
& $10^{11}$ & $0.28\pm0.03$ & $-2.5\pm0.1$ & $2.4$  & $6$  &  & & & & & & & \\
satellites
& $5\times 10^{11}$ & $0.31\pm0.03$ & $-2.6\pm0.1$ & $1.4$ & $7$ & $0.26\pm0.13$ & $-2.5\pm0.3$ & $0.09$ & $3$ & $0.31\pm0.09$ & $-2.5\pm0.2$ & $0.20$ & $4$\\
(single power law)
& $10^{12}$ & $0.35\pm0.03$ & $-2.7\pm0.1$ & 3.2 & 8 & $0.29\pm0.03$ & $-2.6\pm0.1$ & 5.6 & 6 & $0.27\pm0.03$ & $-2.6\pm0.1$ & 2.3 & 6 \\
& $2\times 10^{12}$ & $0.35\pm0.03$ & $-2.7\pm0.1$ & 1.4 & 8 & $0.30\pm0.04$ & $-2.6\pm0.2$ & 0.23 & 6 & $0.31\pm0.02$ & $-2.6\pm0.1$ & 0.12 & 6 \\
& $10^{13}$ & & & & & $0.39\pm0.03$ & $-2.9\pm0.1$ & 4.2 & 8 & $0.38\pm0.03$ & $-2.9\pm0.2$ & 2.4 & 8 \\
& $10^{14}$ & & & & & $0.39\pm0.04$ & $-3.0\pm0.2$ & 1.4 & 8 & $0.41\pm0.03$ & $-3.0\pm0.2$ & 2.2 & 8 \\
& $10^{15}$ & & & & & $0.41\pm0.03$ & $-3.1\pm0.2$ & 2.0 & 8 & $0.44\pm0.03$ & $-3.2\pm0.1$ & 1.3 & 8 \\
\hline
& $M_*/\msun$  & $\alpha$ & $b$ & $\chi^2$ & $N_{\rm bin}$ & $\alpha$ & $b$ & $\chi^2$ & $N_{\rm bin}$ & \multicolumn{4}{c}{}  \\
\hline
{\rm central galaxies}  & $10^3$--$10^{11}$ & $0.30\pm0.01$ & $-2.52\pm 0.05$ & 18.8 & 16 & & & & & \multicolumn{4}{c}{} \\
{\rm (single power law)}  & & & & & & & & & & \multicolumn{4}{c}{} \\
\hline

& $10^3$--$10^{8}$ ($10^{13}$)\footnote{\tiny The values in the brackets are the host halo masses in unit of $h^{-1}\msun$.} & & & & & $0.31\pm0.05$ & $-2.7\pm0.2$ &  &  &  &  &  &  \\
& $10^{8}$--$10^{11}$ ($10^{13}$) & & & & & $0.54\pm0.08$ & $-3.8\pm0.5$ & 0.4\footnote{\tiny The total $\chi^2$ in the mass range $10^3$--$10^{11}\msun$.} & 8\footnote{\tiny The
total bin number in the mass range $10^3$--$10^{11}\msun$.} &  &  &  &  \\

satellites
& $10^3$--$10^{8}$ ($10^{14}$) & & & & & $0.34\pm0.06$ & $-2.8\pm0.2$ &  &  &  &  &  &  \\
(double power law)
& $10^{8}$--$10^{11}$ ($10^{14}$) & & & & & $0.50\pm0.09$ & $-3.6\pm0.6$ & 0.04 & 8 &  &  &  &  \\

& $10^3$--$10^{8}$ ($10^{15}$) & & & & & $0.35\pm0.06$ & $-2.9\pm0.2$ &  &  &  &  &  &  \\
& $10^{8}$--$10^{11}$ ($10^{15}$) & & & & & $0.53\pm0.09$ & $-3.8\pm0.5$ & 0.1 & 8 &  &  &  &  \\
\hline
{\rm central galaxies} & $10^3$--$10^{8}$ & $0.24\pm0.02$ & $-2.38\pm 0.06$ & & & \multicolumn{8}{c}{} \\
{\rm (double power law)} & $10^{8}$--$10^{11}$ & $0.47\pm0.02$ & $-3.5\pm 0.1$ &  &  & \multicolumn{8}{c}{} \\
              & $>10^{11}$ & 0 & $0.26\pm 0.01$ & 8.3\footnote{\tiny The total $\chi^2$ over the whole mass range ($>10^3\msun$).}  & 20\footnote{\tiny The total bin number over the
whole mass range ($>10^3\msun$).} & \multicolumn{8}{c}{} \\
\hline
\hline
{\rm observation} & & \multicolumn{12}{c}{$\alpha=0.30\pm0.02$,  $b=-2.59\pm 0.04$} \\
\hline
\end{tabular}
}
\tablecomments{The fitting results to the MZR of the satellites and the central
galaxies obtained by the simulation runnings listed in Table~\ref{tab:runs},
which are labeled by ``satellites'' and ``central galaxies'', respectively. In
the table,  $\alpha$ is the best-fit slope, $b$ is the best-fit intercept value
of [Fe/H] at $\log(M_*/\msun)=3$ (see Equation~\ref{eq:MZR}), and $N$ is the
number of the stacked halos in the fitting.  In the table, we show the results
obtained by the two fitting ways for the MZR: (1) one linear slope fitting in
the stellar mass range $10^3\msun<M_*<10^{11}\msun$; and (2) a continuous
double power law fitting with a break at $M_*=10^8\msun$ in the stellar mass range
$10^3\msun<M_*<10^{11}\msun$. There is also a constant fitting at
$M_*>10^{11}\msun$ in the double power law fitting to the central galaxies.  The
$\chi^2$ is the least-squares value of the fitting, and $N_{\rm bin}$ is the
number of the bins used in the fitting.  The one linear slope fitting has a
degree of freedom $N_{\rm bin}-2$, and the second continuous fitting with a
double slope has a degree of freedom $N_{\rm bin}-3$.  The fitting results of
the simulated satellites are listed in order of their host halo masses at
$z=0$, and their single power law fitting results are also shown by the solid dots in
Figure~\ref{fig:mzr_ab}(c)-(d).  The simulated central galaxies
with a large range of host halo masses cover a large range of stellar masses,
as shown in Figure~\ref{fig:cet_full}.
The last row represents the observational result
of the Local Group dwarfs (K13), which is labeled by ``observation''.  See more
details in Section~\ref{sec:res}.
} \label{tab:stack} 
\end{table*}

In this subsection, we generalize the exploration of the MZR of the satellites
to other halo masses (ranging from $10^{10}h^{-1}\msun$ to
$10^{15}h^{-1}\msun$).  Figure~\ref{fig:sats_full} shows the simulated MZRs of
the satellites in the host halo masses ranging from $5\times
10^{11}h^{-1}\msun$ to $10^{15}h^{-1}\msun$, and each panel illustrates the
stacked results of 10 merger trees of the same halo mass.  The 10-tree results
for $M\halo<5\times 10^{11}h^{-1}M_{\odot}$ are not shown in this figure, as
the number of their satellites is statistically small (the statistical results
of their 1000 trees are listed in Table~\ref{tab:stack} below).  As seen from
this figure, the correlation between stellar metallicities and stellar masses of
the satellites exists for the different halo masses. 

\begin{figure*}[htb!] \centering
\subfigure{\includegraphics[scale=0.73]{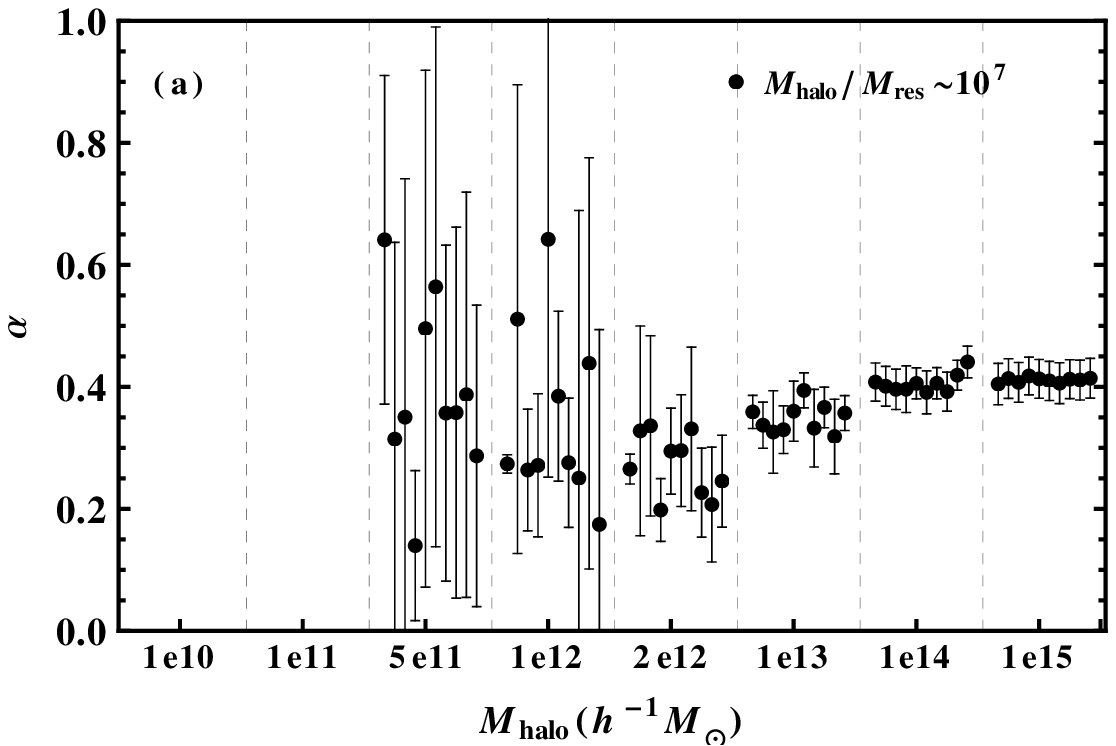}}
\subfigure{\includegraphics[scale=0.73]{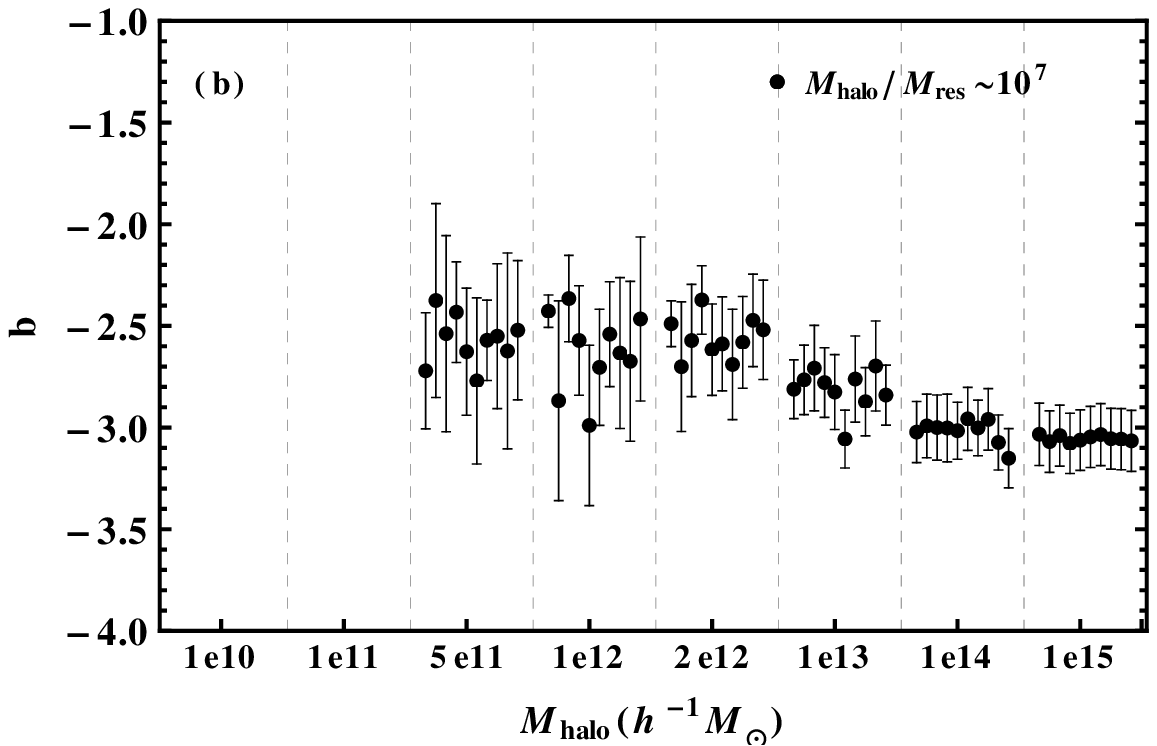}}
\subfigure{\includegraphics[scale=0.73]{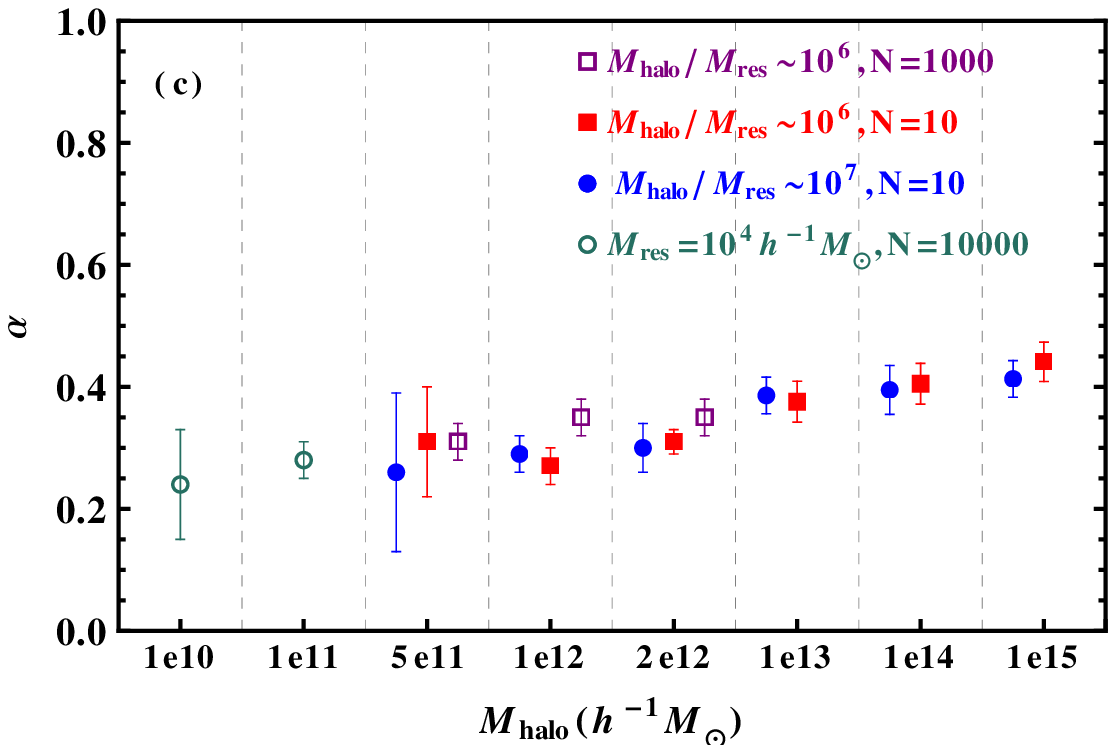}}
\subfigure{\includegraphics[scale=0.73]{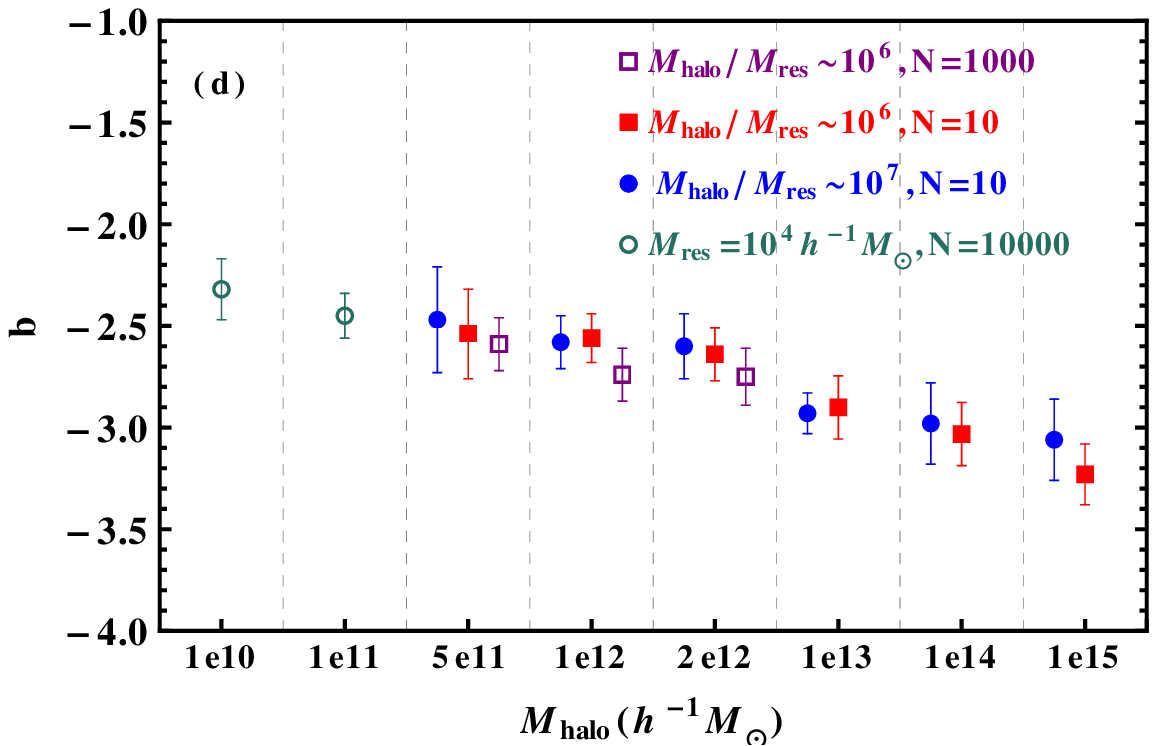}}
\subfigure{\includegraphics[scale=0.73]{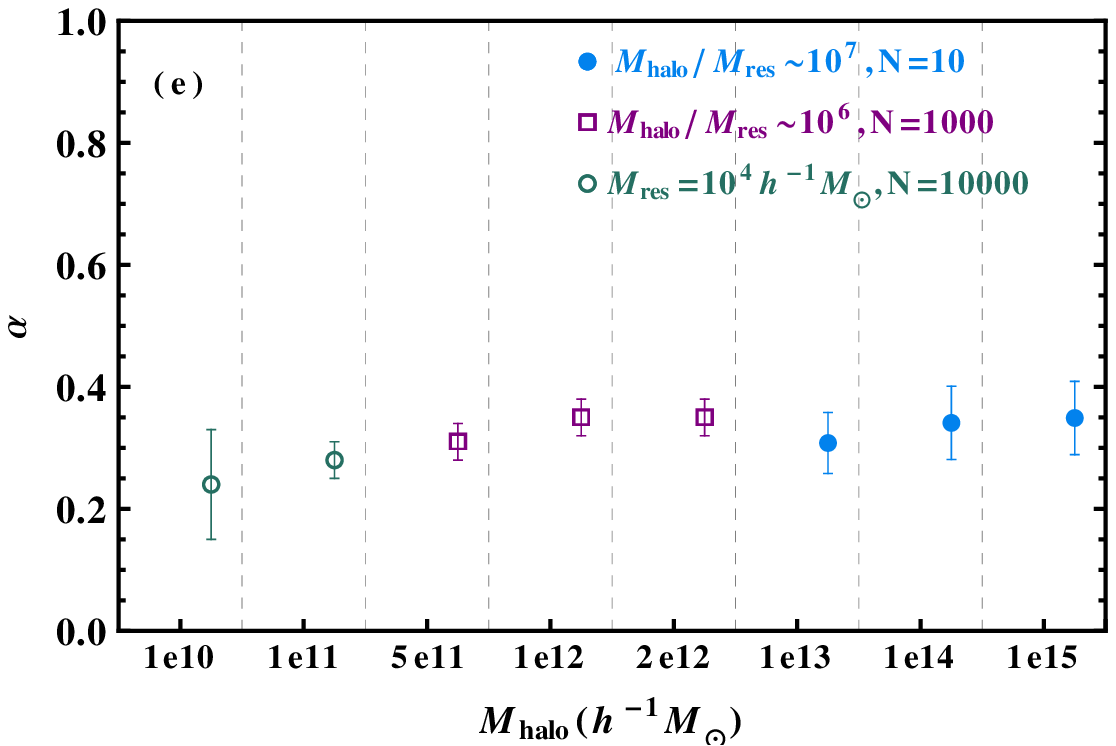}}
\subfigure{\includegraphics[scale=0.73]{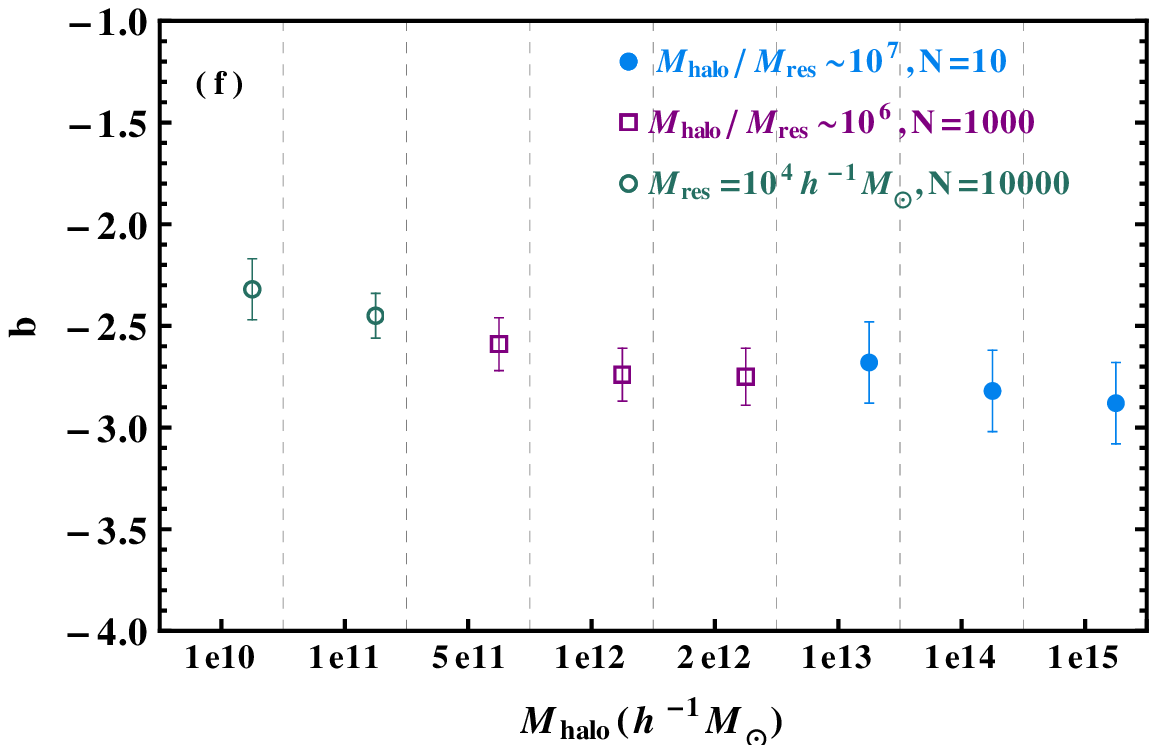}}
  \caption{The fitting results of the MZR of the simulated
satellites in dark matter halos with different masses. Panels (a)-(d) show the linear
(single power law) fitting results, and panels (e)-(f) include the two-slope
fitting results. The left panels present the best-fit
slopes, and the right panels present the best-fit intercepts at
$\log(M_*/\msun)=3$.  In each panel, the results of different host halo masses
are separated by the vertical dashed lines, the points within two adjacent
vertical lines have the same host halo masses exactly as labeled on the
horizontal axis.  Within two adjacent vertical lines, we do not overlap the
points to be in the same  position, but shift them leftwards or
rightwards each other, for clarity.  In panels (a)-(b), each point
represents the result of one individual tree, and in panels (c)-(f), each
point represents the result of a stack of $N$ trees.  The results shown in
panels (c)-(f) are also listed in Table~\ref{tab:stack}.  The detailed fitting
method is described in Sections~\ref{subsec:MWlike}--\ref{subsec:multihalo}.
Panels (a)-(b) show that the fitting results obtained from the individual
merger trees with the same halo masses do not differ significantly for
$M\halo\ga 2\times 10^{12}h^{-1}\msun$, but can scatter significantly in halos
with lower masses.  In panels (c)-(d), we show that the results are not
affected much by the settings of the  mass resolutions in the simulations, as
the results obtained with two halo mass resolutions $M\res$ are close to being
convergent (see the solid points with $N=10$).  The slopes obtained with $N=10$
(solid squares) are a little lower than those obtained with $N=1000$ (open
squares) in low-mass halos, because there are relatively few high-mass
satellites obtained with the small number of $N$ in the fitting. 
%
%
Panels (c)-(d) show that the MZR of the stacked satellites is roughly universal with
different halo masses, with the slopes being in the same range $\sim$ 0.2--0.4
(though slowly increasing with increasing halo masses) and with intercepts
being in the range $\sim-2.3$--$-3.1$ at $\log(M_*/\msun)=3$.
In panels (e)-(f), the points shown for $M\halo=10^{10}$--$2\times 10^{12}h^{-1}\msun$
are the same as those points with the same types shown in panels (c)-(d), and the points
shown for $M\halo=10^{13}$--$10^{15}h^{-1}\msun$ are the double power law fitting results
at stellar mass $M_*=10^3$--$10^8\msun$ as listed in Table~\ref{tab:stack}.
As described in Section~\ref{subsec:multihalo}, the
double power law fitting to the MZR in high-mass halos results in a relatively low
$\alpha$
for low-mass satellites and a high $\alpha$ for high-mass satellites, which yields a better
fit and adds the support to the universality in the MZRs of the satellites.
In addition, the slight decrease of the slope with decreasing halo masses in
low-mass halos with $M_{\rm halo}$=$10^{10}$--$10^{12} h^{-1}\msun$ shown in
panel (e) is associated with the monotonically increasing intercept $b$ in the
same halo mass range shown in panel (f) (see also Figure~4c or 4d below), which
is related to the metallicity enrichment caused by SNe Ia, as well as different
halo assembly histories with different halo masses and their different star
formation histories at high redshifts (see discussion in Section 3.2).
}\label{fig:mzr_ab}

\end{figure*}

As described in Section~\ref{subsec:MWlike}, we use the linear least-squares
method to fit the MZR by Equation (\ref{eq:MZR}) for each tree shown in
Figure~\ref{fig:sats_full}.  Note that the stellar metallicities start to
become flat at the high-mass end ($M_*\ga 10^{11}\msun$). The linear fitting is
limited to the range $10^3\msun< M_*<10^{11}\msun$. The best-fit parameters
obtained for the satellites in 10 individual halos are shown in 
Figure~\ref{fig:mzr_ab}(a)-(b).  For high halo masses, the best-fit
parameters of the individual halos do not scatter significantly. In low-mass
halos, the small number statistics of the satellites in one halo can affect the
best-fit parameters significantly. To view the results in a statistical way, we
stack the satellites of the 10 halos together and get the best-fit results, and
we show them in panels (c) and (d) of the figure. Panels (c) and (d) also include
the best-fit results to the stacked 10000 or 1000 halos with low halo masses.
Table~\ref{tab:stack} summarizes the MZR fitting results of the stacked halos.

According to the linear (single power law) fitting results shown in
Figure~\ref{fig:mzr_ab}(c)-(d) or listed in Table~\ref{tab:stack}, we find the
following results.
\begin{itemize}
\item Our fitting results are not affected much by the mass resolution set in
the simulations, because the results obtained with the two different halo
resolution masses $M\res$ are close to being convergent for the same halo
masses with the same stacked halo number $N=10$;
\item The MZR slopes appear to be relatively lower in low-mass halos, compared
to the slopes in high-mass halos, mainly because the low-mass halos have few
high-mass satellites and those high-mass satellites play an important role in
the fitting. 
For the halo masses that are not very high (e.g., $M\halo=5\times 10^{11}$ or
$10^{12}h^{-1}\msun$), an increase of the number of the stacked halos (e.g.,
from $N=10$ to $N=1000$) can increase the fitted MZR slopes, as the number of
high-mass satellites increases. The increase of the slopes in high-mass halos
should be taken to be caused by a physical effect in high-mass galaxies, as a
double power law fitting to the MZR yields a higher slope in high-mass galaxies for
both central galaxies and satellites in Section~\ref{subsec:central} and at
the end of this subsection.
\end{itemize}
Figure~\ref{fig:mzr_ab}(c)-(d) and Table~\ref{tab:stack} show that for the MZRs of the
stacked satellites in different host halo masses, the average MZR slopes obtained
by the linear (single power law) fit are
generally in the same range $\sim$0.2--0.4 and slowly increase with increasing
halo masses, and the average MZR intercept $b$ at $\log(M_*/\msun)=3$ are
generally in the range $\sim -2.3$ to $-3.1$.

The $\chi^2$ values shown in Table~\ref{tab:stack} are used as a measure of the
goodness of the fit, together with the number of the bins $N_{\rm bin}$ in the
fit.  In general the linear (single power law) fits are acceptable, and the
probability that a random set of $N_{\rm bin}$ data points drawn
from the parent distribution would yield a value of $\chi^2$ as large as or
larger than the tabulated values are mostly in the range $\sim50\%-95\%$. 
However, we have the following notes on the use and the limitation of the
results.
\begin{itemize}
\item As mentioned in Section~\ref{subsec:MWlike}, we do not pre-assume that
the stellar metallicities follow a Gaussian distribution at a given stellar
mass.  For the scatters of data points not following an exact Gaussian
distribution, the fit parameters and their errors are not the ``minimum
variance unbiased estimator'' (MVUE) in the least-squares statistics, but the
least-squares method is still helpful by being generalized to compare data with
models (e.g., see Chapters 4, 6, and 11 in Bevington \& Robinson 2003), which
serves as a ``first-order'' approximation to understand the data
quantitatively, as well as a mathematical way to compare the different models
as done in this work. 
\item Some $\chi^2$ values are somewhat too small, compared with
the degrees of freedom in the fitting (related with the bin number as described
in the note of Table~\ref{tab:stack}). Part of the reason for this is
that a relatively large error has been assigned to the fitting data (i.e.,
using the large scatter of the [Fe/H] around the median in each bin).
\item Understanding the fit parameters and their errors better will involve
understanding the underlying physical processes/reasons leading to the data,
which should be taken into account in the comparison of observational data with
simulation results.
\end{itemize}

As the slopes obtained by a linear fitting to the MZR slightly increases with
increasing halo masses within the range $\sim0.2$--0.4, we perform a
double power law fit for the satellites in the high-mass halos
($M\halo=10^{13},10^{14},10^{15} h^{-1}\msun$), with a break at $M_*=10^8\msun$
in the stellar mass range $10^3\msun<M_*<10^{11}\msun$.  The fitting lines are
continuous at the break point.  The fitting is done by the least-squares method
as described in Section~\ref{subsec:MWlike}. The fitting results are shown in
Table~\ref{tab:stack} and Figure~\ref{fig:sats_full}(d)-(f).
 We do not perform a double power law fit for low-mass
halos, as their satellite masses are not high enough.  As seen from the
double power law fitting results shown in Table~\ref{tab:stack}, we find that
the double power law fitting yields a much smaller $\chi^2$ value, which
indicates a better fit than the single power law fitting (though
the $\chi^2$ values are too small compared with the degrees of freedom,
which is caused by the large errors assigned to the fitting data, as mentioned
above).  In the low-mass range with $10^3\msun<M_*<10^{8}\msun$,
the fit slopes ($\sim 0.31$--$0.35$) become smaller than their single power law fitting
results and closer to the linear fit slopes in the low-mass halos; and the
difference of the intercepts from those in low-mass halos also becomes smaller,
which is $\la0.6$~dex at $\log(M_*/\msun)=3$ or $\la0.2$~dex at
$\log(M_*/\msun)=6$.  In the high-mass range with $10^8\msun<M_*<10^{11}\msun$,
the slopes obtained by the double power law fitting is roughly universal, $\sim
0.5$, which are higher than their single power law fitting results. The results of the
double power law fitting to high-mass halos supports that there is no significant
difference in the MZRs of the satellites in host halo masses $\sim
10^{11}$--$10^{15}h^{-1}\msun$, though the MZR slopes in lower halo masses
are still slightly smaller.  The results of the
double power law fitting support a universal MZR in satellites. 

The slight decrease of the slope with decreasing halo masses existing at the
low-halo mass range of $M_{\rm halo}\sim 10^{10}$--$10^{12} h^{-1}\msun$ is
associated with the monotonically increasing intercept $b$ in the same halo
mass range shown in Figure 3(f) (or Figs.~4c and 4d below), which is related to
the metallicity enrichment caused by SNe Ia, as well as different halo assembly
histories with different halo masses and their different star formation
histories at high redshifts.  As discussed in section 3.1.1 in HYL14, the
scatter of the simulated MZRs is generally large at the low-$M_*$ end, which is
caused mainly through the scatter in the star formation durations of the
galaxies at a given stellar mass and the difference in the chemical enrichment
of SNe Ia and II. If the duration is short, the metal enrichment is mainly
contributed by SN II explosions, which have a chemical pattern with a
relatively low iron fraction; if the duration is long enough, SNe Ia may have a
non-negligible contribution to the metal enrichment and generate more iron than
SNe II. Thus, a short star formation duration would lead to a low [Fe/H], while
a longer one leads to a higher [Fe/H]. High-mass satellites generally all
experienced extended star formation duration; low-mass satellites are generally
formed at high redshifts and some of them had relatively shorter star formation
duration. The low-mass satellites with relatively shorter star formation
duration have relatively lower metallicities, and this effect appears to be
slightly more significant in high-mass halos. We have also done the test to
obtain the MZR results by removing the metallicity enrichment due to SN Ia in
the semianalytical galaxy formation and evolution model and find that the
slight decrease of the slopes and the increase of the intercepts with
decreasing halo masses in low-mass halos with $M_{\rm halo}\sim
10^{10}$--$10^{12} h^{-1}\msun$ disappear (where the intercepts of the
satellite MZRs in low-mass halos become $\sim-2.7$ to $-2.9$).

\subsection{The MZR of central host galaxies}\label{subsec:central}

\begin{figure*}[htb] 
\centering
  \subfigure{\includegraphics[scale=0.7]{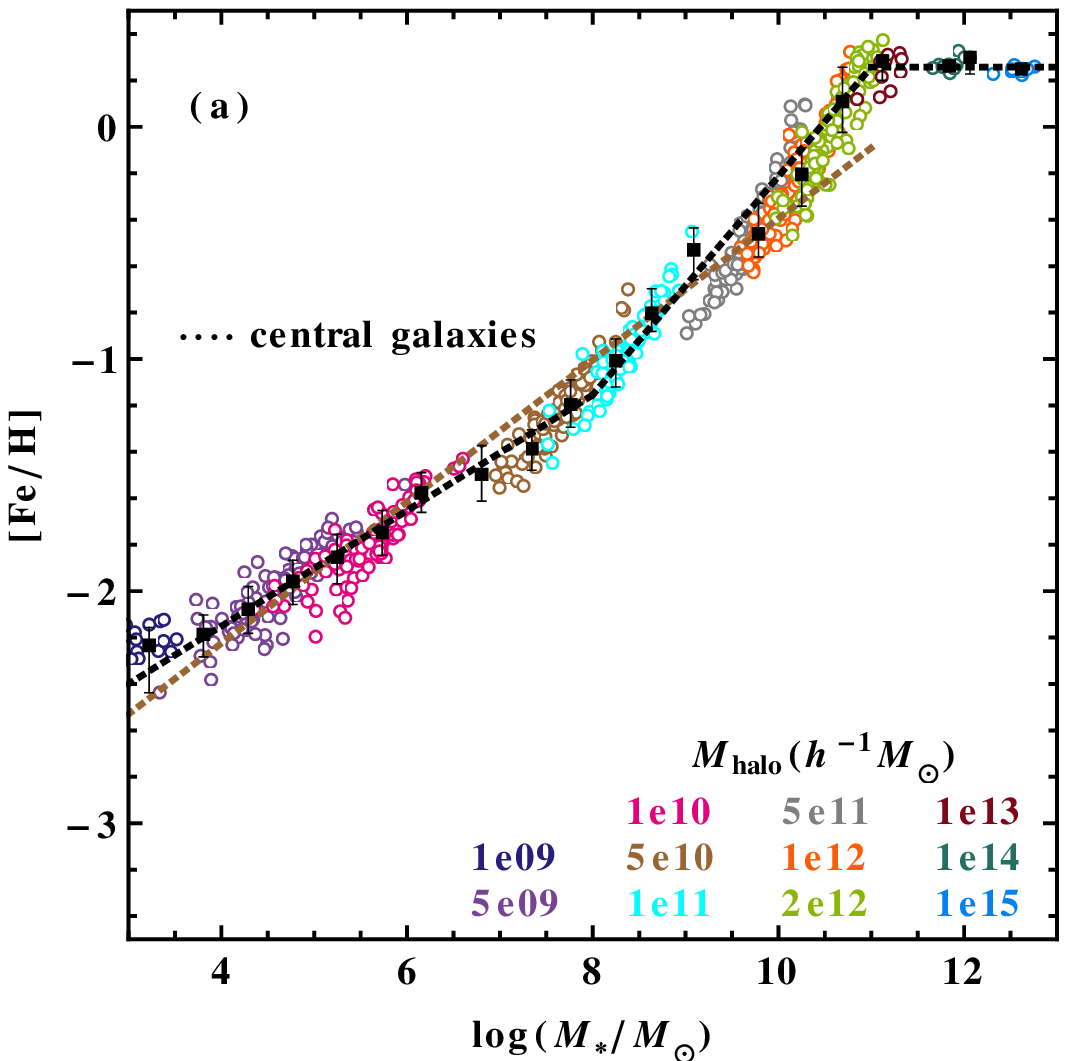}}
  \subfigure{\includegraphics[scale=0.7]{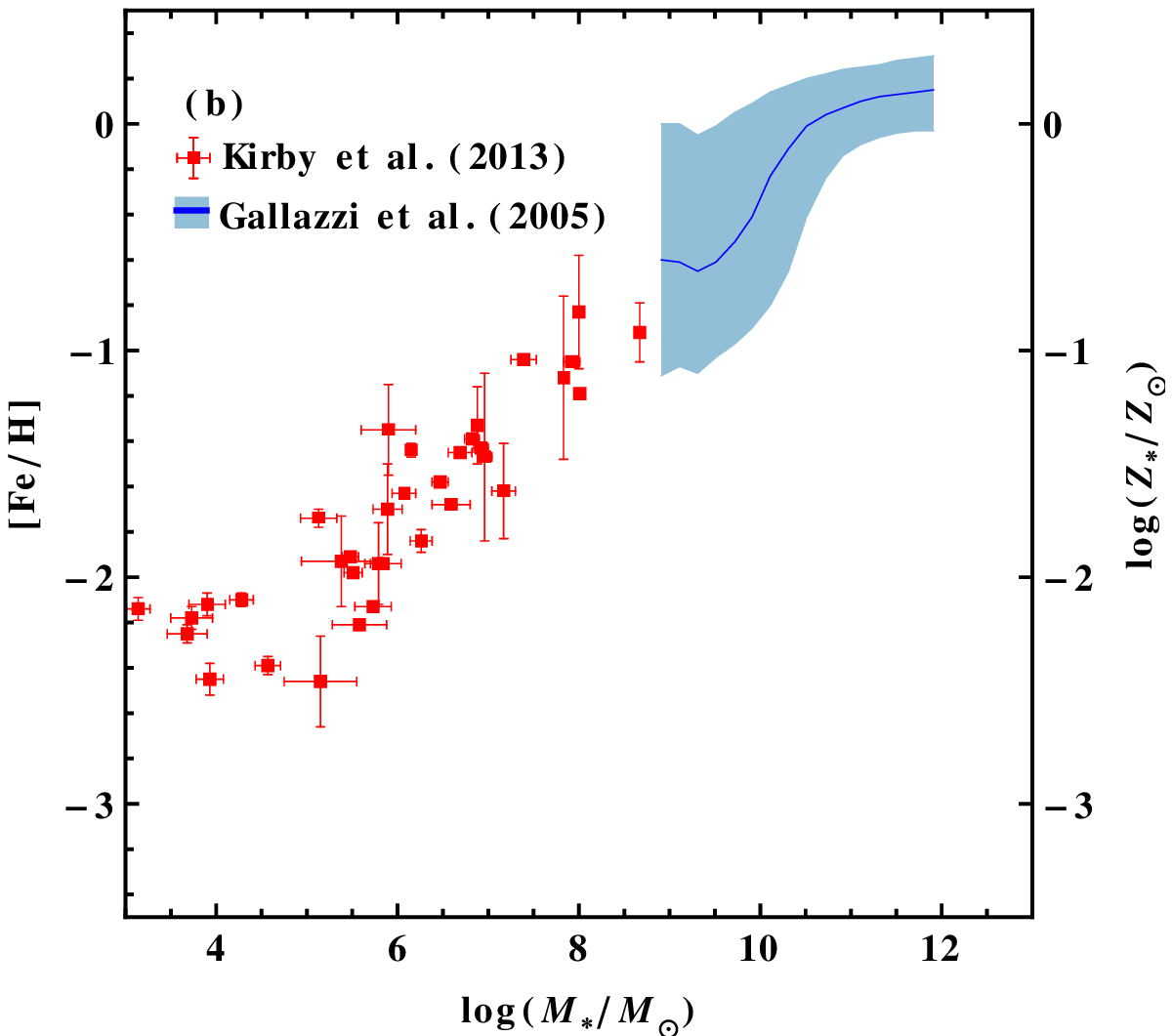}}
  \subfigure{\includegraphics[scale=0.7]{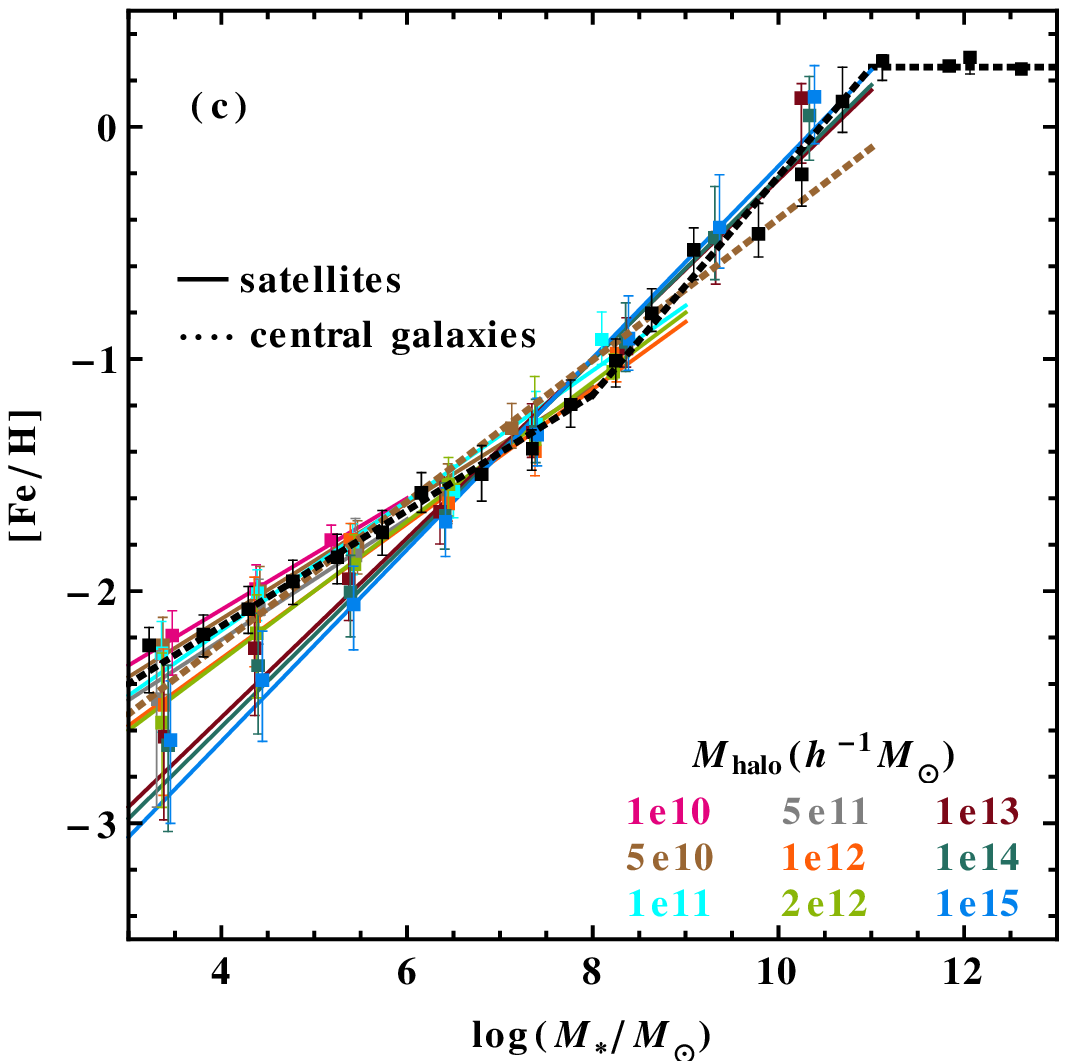}}
  \subfigure{\includegraphics[scale=0.7]{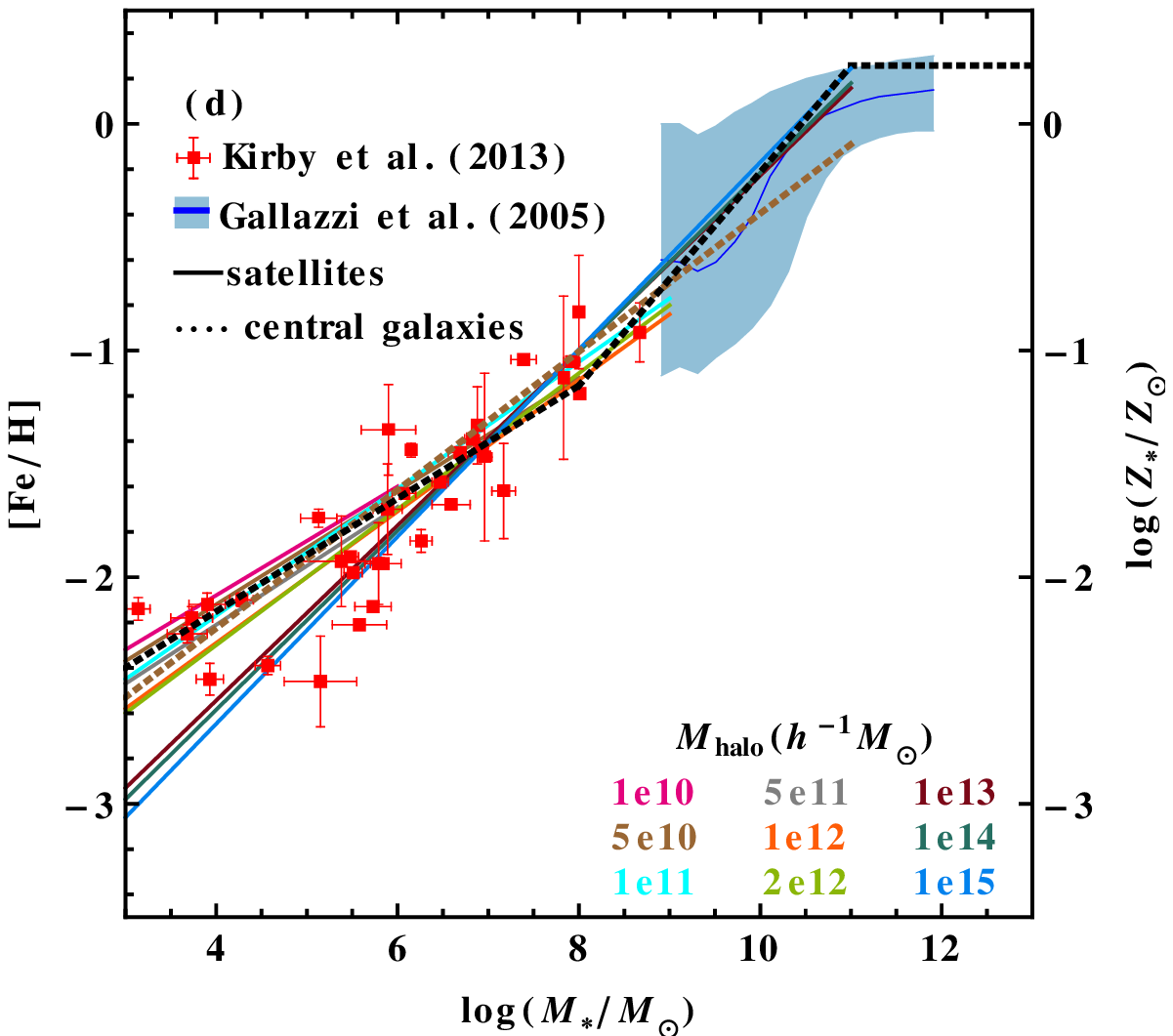}}
  \caption{The MZR of central host galaxies obtained from observations and our
simulations. (a) The MZR of the simulated central galaxies. The results are
obtained from the simulation runnings in the left column of
Table~\ref{tab:runs}. Each circle represents one simulated central galaxy.
Different colors represent different host halo masses, as indicated by the
text. For view clarity, we show only 100 points randomly selected from the 10000
or 1000 runnings for each halo mass in the range of $10^9$--$2\times
10^{12}h^{-1}\msun$.  The brown dotted line is the linear fitting result to the
simulation data at $10^3\msun<\log M_*<10^{11}\msun$.  The black dotted line is
the double power law fitting result to the simulation data at $10^3\msun<\log
M_*<10^{11}\msun$ and a constant fitting to the simulation data at $\log M_*>
10^{11}\msun$, where the values of [Fe/H] are fit to be continuous at the break
points. As shown by the black dotted line, the MZR at $10^8\msun \la \log
M_*\la 10^{11}\msun$ has a steeper slope than that at $10^3\msun \la \log
M_*\la 10^8\msun$.  The best-fit parameters are listed in Table 2.
(b) The red filled squares are the observational results for dwarf galaxies in
the Local Group (see Figure 9 in K13).  The blue solid line gives the median of
the MZR for 44,254 late-type galaxies drawn from SDSS DR2 \citep{Gallazzietal05},
and the light blue region is between the 16th and 84th percentiles of the
distribution.  Note that the metallicities in the Local Group dwarfs and in the
SDSS late-type galaxies are measured through different techniques.  The stellar
metallicities of the Local Group dwarfs are expressed through [Fe/H] labeled on
the left side of the panel, while the stellar metallicities of the SDSS
late-type galaxies are expressed through $\log(Z_*/Z_\odot)$ labeled on the
right side of the panel.  As mentioned in K13, the conversion between [Fe/H]
and $\log Z_*$ depends on [Mg/Fe] or the abundance ratio of alpha elements to
Fe.  (c) Comparison between the fitting results of the simulated central
galaxies and the simulated satellites.  The dotted lines are the same as those in
panel (a).  The solid lines (together with their square points), the same
ones as those in Figure~\ref{fig:sats_full}, represent the linear fitting results
of the satellites in one halo mass, and different
colors represent the different host halo masses, as indicated in the text. 
Only the single power law fitting results are illustrated in this panel, because
the linear fitting is normally taken as the initial attempt to analyze the
data (with a limited size) obtained from observations and the fitting results
serve as a ``first-order'' approximation to understand the data quantitatively.
In the panel, the high-mass ends of those solid lines for the satellites
in halo masses $M\halo\le 2\times 10^{12} h^{-1}\msun$ are extended only to the
highest-mass bin used in the fitting, not to $M_*=10^{11}\msun$.
(d) Combination of the results in (b) and (c).  The figure shows that our
simulation results can reproduce the observations. The simulated
MZRs in satellites with different halo masses and in
central galaxies are roughly universal, with their slopes
and intercepts distributed within a relatively narrow range. }\label{fig:cet_full}
\end{figure*}

We obtain the MZR for the simulated central galaxies with different host halo
masses. Our model results and the comparison with observations are shown in
Figure~\ref{fig:cet_full}. Figure~\ref{fig:cet_full}(a) shows the MZR of the
simulated central galaxies in different host halo masses, where the stellar
metallicity increases with increasing stellar masses and becomes flat at $\log
(M_*/\msun)\ga 11$. The dotted lines are the fitting results to the data
obtained from the simulation runnings in the left column of
Table~\ref{tab:runs}. The fitting is performed in these two ways: one is fit by
one linear slope within the stellar mass range $10^3\msun<M_*<10^{11}\msun$
(brown dotted line), and the other is fit by two slopes with a break at
$M_*=10^8\msun$ in the stellar mass range $10^3\msun<M_*<10^{11}\msun$ and a
constant at $M_*>10^{11}\msun$ (black dotted line), where the fitting lines
are continuous at the break points. The fitting is done by the
least-squares method as described in Section~\ref{subsec:MWlike}. A relatively
small bin size with 0.5~dex in $\log M_*$ (compared to 1-dex interval used 
for the satellites above) is used due to relatively more
high-mass systems here. The best-fit slopes and
intercepts are listed in Table~\ref{tab:stack}.  As seen from
Table~\ref{tab:stack}, the single power law fit gives the MZR slope $\alpha=0.30\pm
0.01$ with $b=-2.52\pm0.05$, which is in the same range as the MZR slopes of their
satellites. The double power law fit gives a steeper slope ($\alpha=0.47\pm 0.02$)
in the mass range $10^8\msun<M_*<10^{11}\msun$ than that ($\alpha=0.24\pm
0.02$) in the relatively low-mass range $10^3\msun<M_*<10^{8}\msun$.
This tendency is consistent with the above MZR fitting results for satellites
described in Section~\ref{subsec:multihalo}.  The slope of the central galaxies
in the low-$M_*$ range ($\alpha=0.24\pm 0.02$) is at the low end of the slope
range of the satellites ($\alpha\sim0.2$--0.4), indicating that many low-mass
satellites are relatively more metal poor.  In the high-mass range with
$10^8\msun<M_*<10^{11}\msun$, the best-fit slope is roughly the same as the
corresponding values of the satellites obtained in the double power law fit.
The metallicity becomes roughly constant at $M_*\ga 10^{11}\msun$, with
[Fe/H]$\sim 0.26$, which is mainly determined by the Fe yield of SNe II and Ia.
As seen from Table~\ref{tab:stack}, the $\chi^2$ value of the double power law fit
for the central galaxies are smaller than that of the
single power law fit, which suggests that the double power law fit is acceptable with a
higher probability.
Figure~\ref{fig:cet_full}(b) shows the observational results of some Local
Group dwarfs and nearby galaxies (K13; Gallazzi et al.\ 2005).
Figure~\ref{fig:cet_full}(c) shows that 
the MZRs of the simulated central galaxies (dotted line) and the simulated
satellites (solid lines; same as those in Figure~\ref{fig:sats_full}) are
universally distributed within the same space.
Figure~\ref{fig:cet_full}(d) shows that our simulation results of the MZR
generally agree well with the observational data.

When more observational results with good measurement quality come out and
support the nonlinear behavior in
the MZR or the double power law fitting does not perform well, it would be helpful
to perform a nonparametric or semi-parametric regression, which requires large
sample sizes, on both observational data and simulation data to uncover the
underlying data structure and the contributions from different physical
processes or high-order physical effects (e.g., see some
modern statistical methods in \citealt{FB12,T05,RWC03}).

\subsection{The MZR in Local Group dwarfs}

Our simulation results obtained above support that Local Group dwarfs follow a
universal MZR at least by the following points. (1) As shown in
Figure~\ref{fig:mw100}, the MZR in the satellites of MW-size halos is
universal. M31 is another big galaxy in the Local Group, and the host halo mass
of M31 is also possibly in the range $\sim 1$--$2\times 10^{12}\msun$ (e.g.,
\citealt{Kafleetal18} and references therein). The results shown in
Figures~\ref{fig:mw100} and \ref{fig:sats_full} support that the MZR of the
satellites in both M31 and MW follows a universal MZR.  (2)
Figure~\ref{fig:cet_full} shows that the central galaxies of smaller halos with
$M\halo<10^{11}h^{-1}\msun$ follow the same universal MZR as that for
satellites in the MW-size halos.  Other dwarf galaxies in the Local Group can
be explained as the central galaxies of smaller halos, so they also follow the
universal MZR. 

\section{Summary and Discussion}

We have investigated the stellar mass--stellar metallicity relations of
galaxies and their satellites by using the semi-analytical models. Our study
suggests a universal correlation of the stellar metallicity with stellar
mass $M_*\sim 10^3\msun$--$10^{11}\msun$, [Fe/H]=$\alpha\log (M_*/10^3\msun)+b$,
and a roughly constant stellar metallicity for $M_*\ga 10^{11}\msun$.  The
relations reproduced from our work are consistent with the observations on the
MW/M31 dwarf satellites ($M_*\sim 10^3$--$10^{8.5}\msun$) and local galaxies
($M_*\sim 10^9\msun$--$10^{12}\msun$), as shown in Figure~\ref{fig:cet_full}.

Our study shows that the slope $\alpha$ of the MZR for the satellites in a
halo with mass the same as MW/M31 halo mass ([Fe/H]--$\log M_*$) are mostly in
the same range of 0.2--0.4 (i.e., with a scatter of $\sim 0.2$ for different
halo assembly histories), and the intercept at $\log(M_*/\msun)=3$ is
$b\sim -2.5$ to $-2.9$.
If the satellites of many halos are stacked together, the MZR for the stacked
satellites in MW/M31-size halos gives $\alpha \simeq 0.35\pm0.03$ and the
intercept $b\simeq -2.7\pm 0.1$.  The slopes of the stacked satellites in the
host halo masses ranging from $10^{10}h^{-1}\msun$ to $10^{15}h^{-1}\msun$ are
correspondingly in the same range of 0.2--0.4, with a slight increase with
increasing halo mass; and the intercept $b$ at $\log(M_*/\msun)=3$ is in the
range $-2.3$ to $-3.1$.  

Our study shows that the slope of the correlation for the central galaxies with
$M_*$ from $\sim 10^3$ to $10^{11}\msun$ is $\alpha\simeq 0.30\pm 0.01$ with
$b=-2.52\pm0.05$, which is in the same range of the satellite MZRs. 
The MZR becomes roughly constant at $M_*\ga 10^{11}\msun$, with
[Fe/H]$\sim 0.26$, which is mainly determined by the Fe yield of SNe II and Ia.

Our study also shows that a double power law provides a better fit to the
stellar mass--stellar metallicity relation than the single power law for both
satellites and central galaxies. The double power law fit gives
$\alpha\sim$0.2--0.4 at $10^3\msun\la M_*\la10^{8}\msun$ and a relatively
higher $\alpha\sim0.5$ at $10^8\msun\la M_*\la10^{11}\msun$. The difference in
the best-fit intercepts of the correlations is $\la 0.6$~dex at $\log(M_*/\msun)=3$ (with $b\sim-2.9$ to $-2.3$) or  $\la0.2$~dex at $\log(M_*/\msun)=6$.
Specifically, the double power law fit to the central galaxies gives
$\alpha\simeq 0.24\pm 0.02$ (close to the low end of the range 0.2--0.4) and
$b\simeq -2.38\pm 0.06$ at $10^3\msun<M_*<10^8\msun$, and $\alpha\simeq 0.47\pm
0.02$ and $b\simeq -3.5\pm 0.1$ at $10^8\msun<M_*<10^{11}\msun$.  The high-mass
satellites ($M_*\ga10^8\msun$) existing mostly in high-mass halos and their
relatively high $\alpha$ can lead to an apparent increase of $\alpha$ with
increasing host halo masses obtained in the single power law fitting to the
satellites; and after taking into account that effect, the dependence of the
satellite MZRs on their host halo masses becomes little in the halo mass range
$M\halo\sim 10^{11}$--$10^{15}h^{-1}\msun$, though the slopes in lower
halo masses are still slightly smaller. 

For the MZRs of satellites in low-mass halos with $M_{\rm halo}\sim
10^{10}$--$10^{12} h^{-1}\msun$, the slight decrease of the slope with
decreasing halo masses is associated with their monotonically increasing
intercept $b$ with decreasing halo masses, which is related to the metallicity
enrichment caused by SNe Ia, as well as different halo assembly histories with
different halo masses and their different star formation histories at high
redshifts.

Although a detailed examination of the simulated MZRs may suggest some slight
differences among the central galaxies and their satellites with host halo
masses spanning a wide range ($M\halo\sim 10^{9}$--$10^{15}h^{-1}\msun$), we
call the MZRs ``universal'' because their slopes and normalizations are within
a relatively narrow range.  The universal relation of the MZRs in the
satellites within the large range of halo masses and their central galaxies
awaits future observational tests. A precise comparison with
observational results would require the consideration of observational
selection effects and similar fitting methods.  A double power law or non-linear
fitting to the MZRs would become necessary if the number of the galaxies are
sufficiently large and their masses cover a large range.

Active galactic nuclei (AGN) feedback is not included in the model. AGN feedback can have significant
effects mainly in high-mass galaxies (e.g., $\ga 10^{11}\msun$;
\citealt{Croton06,Boweretal06,SD15}), which decreases the galaxy mass function
at the high-mass end.  As shown in this work, the stellar metallicity increases
with increasing stellar mass mainly in galaxies with $M_*\la 10^{11}\msun$, and
the stellar metallicity becomes roughly constant at higher masses. AGN feedback
would not have significant effects in the monotonically increasing part of the
stellar metallicity with the stellar mass.

For simplicity, the reionization epoch is set to the same for the different
halo masses in this work so that the effects due to other physical processes
can be isolated. A late reionization epoch may result in relatively low
metallicities at the low-mass end of the MZR and a steeper slope, as 
illustrated for MW satellites in HYL14. Future observations on the slope of the
MZR for dwarfs would provide a constraint on whether or how the reionization
epoch is different in different environments.

A modification to the cooling recipe in semianalytical models can result in a
different luminosity function of galaxies. However, the universality in the MZR
revealed in this paper and its compatibility with observations suggest that the
modification itself should not be the key to the origin of the universality.

The universality of the MZR relations in satellites and host galaxies with
different host halo masses provides insights or constraints on understanding the
galaxy formation and evolution and their chemical evolution.  One next
important step of this work is to understand the origin of the universality of
the MZR relation, investigate its evolution with redshift, connect
it to cold gas phase metallicities revealed in observations (e.g.,
\citealt{tremonti04,mannucci10,Leeetal06}), and explore whether there exists a
possible dependence on a third or more parameters (e.g., star formation rate,
as shown in \citealt{Ellisonetal08,mannucci10,BB17}), where the multivariate
Principal Components Analysis or a nonlinear extension that treats more
variables would be useful.  K13 discussed some possible scenarios for the MZR
relations in the MW/M31 satellites.  \citet{F16} give a review on the
scenarios involving inflows and outflows to explain the MZRs in massive
galaxies. The semianalytical model used in this work provides an efficient way
to isolate the effects of different physical processes and to see the dominant
reason leading to the MZRs. The underlying reasons for the
universality of the MZR obtained from the semianalytical models will be
reported in a different work (in preparation).

We thank the referee, Sandra Faber, Youjun Lu, and Eric Peng for helpful
comments on the manuscript.  This work was supported in part by the National
Natural Science Foundation of China under Nos.\ 11673001, 11273004, 10973001,
11721303, the National Key R\&D Program of China (grant No.\ 2016YFA0400703),
and the Strategic Priority Program of the Chinese Academy of Sciences (grant
No.\ 23040100).

\end{document}